\documentclass{aa}

\pdfpageattr{/Group <</S /Transparency /I true /CS /DeviceRGB>>}

\def\GA{\mathrel{\raisebox{0.13\baselineskip}{\hbox{\rlap{\hbox{\lower0.41\baselineskip\hbox{$\sim$}}}\hbox{$>$}}}}}
\def\LA{\mathrel{\raisebox{0.13\baselineskip}{\hbox{\rlap{\hbox{\lower0.41\baselineskip\hbox{$\sim$}}}\hbox{$<$}}}}}

\def\aj{AJ}% Astronomical Journal
\def\araa{ARA\&A}% Annual Review of Astron and Astrophys
\def\apj{ApJ}% Astrophysical Journal
\def\apjl{ApJ}% Astrophysical Journal, Letters
\def\apjs{ApJS}% Astrophysical Journal, Supplement
\def\ao{Appl.~Opt.}% Applied Optics
\def\aap{A\&A}% Astronomy and Astrophysics
\def\aaps{A\&AS}% Astronomy and Astrophysics, Supplement
\def\mnras{MNRAS}% Monthly Notices of the RAS
\def\pasp{PASP}% Publications of the ASP
\def\nat{Nature}% Nature
\def\procspie{Proc.~SPIE}% Proceedings of the SPIE

\usepackage{graphicx}
\usepackage{txfonts}
\usepackage{amsmath}	% Advanced maths commands
\usepackage{amssymb}	% Extra maths symbols
\usepackage{pdflscape}

\usepackage[dvipsnames]{xcolor}
\usepackage{xfrac}
\usepackage{algorithm2e}
\usepackage{multicol}
\usepackage{multirow}
\usepackage{bigstrut}
\usepackage[normalem]{ulem}

\usepackage[english]{babel}
\usepackage{microtype}

\usepackage{lipsum}

\newcommand{\oiii}{\mbox{O\,{\scshape iii}}}

\usepackage{supertabular}

\usepackage[caption=false]{subfig}
\usepackage{float}

\begin{document} 
\label{firstpage}
\selectlanguage{english}

   \title{The Lensed Lyman-Alpha MUSE Arcs Sample (LLAMAS)}
   \authorrunning{A. Claeyssens et al.}

   \subtitle{I. Characterisation of extended Lyman-alpha haloes and spatial offsets.}

   \author{A. Claeyssens\inst{1},
          J. Richard\inst{1},
          J. Blaizot\inst{1},
          T. Garel\inst{2},
          H. Kusakabe\inst{2},
          R. Bacon\inst{1},
          F. E. Bauer\inst{3, 4, 5},
          L. Guaita\inst{6},
          A. Jeanneau\inst{1},
          D. Lagattuta\inst{7,8},
          F. Leclercq\inst{2},
          M. Maseda\inst{9},
          J. Matthee\inst{10},
          T. Nanayakkara\inst{9},
          R. Pello\inst{11},
          T.T. Thai\inst{11, 12, 13},
          P. Tuan-Anh\inst{12, 13},
          A. Verhamme\inst{2},
          E. Vitte\inst{2,14}
          L. Wisotzki\inst{15}
          }

   \institute{Univ Lyon, Univ Lyon1, Ens de Lyon, CNRS, Centre de Recherche Astrophysique de Lyon UMR5574, F-69230, Saint-Genis-Laval, France, \and Observatoire de Gen\`{e}ve, Universit\'e de Gen\`{e}ve, 51 Chemin de P\'egase, 1290 Versoix, Switzerland , \and Instituto de Astrof{\'{\i}}sica and Centro de Astroingenier{\'{\i}}a, Facultad de F{\'{i}}sica, Pontificia Universidad Cat{\'{o}}lica de Chile, Casilla 306, Santiago 22, Chile, \and Millennium Institute of Astrophysics, Nuncio Monse{\~{n}}or S{\'{o}}tero Sanz 100, Of 104, Providencia, Santiago, Chile, \and Space Science Institute, 4750 Walnut Street, Suite 205, Boulder, Colorado 80301 , \and Departamento de Ciencias Fisicas, Universidad Andres Bello, Fernandez Concha 700, Las Condes, Santiago, Chile, \and Centre for Extragalactic Astronomy, Durham University, South Road, Durham DH1 3LE, UK, \and Institute for Computational Cosmology, Durham University, South Road, Durham DH1 3LE, UK, \and Leiden Observatory, Leiden University, P.O. Box 9513, 2300 RA, Leiden, The Netherlands, \and Department of Physics, ETH Zürich, Wolfgang-Pauli-Strasse 27, 8093 Zürich, Switzerland, \and Aix Marseille Université, CNRS, CNES, LAM (Laboratoire d’Astrophysique de Marseille), UMR 7326, 13388, Marseille, France, \and Department of Astrophysics, Vietnam National Space Center, VAST, 18 Hoang Quoc Viet, Hanoi, Vietnam, \and Graduate University of Science and Technology, Vietnam Academy of Science and Technology, 18 Hoang Quoc Viet, Cau Giay, \and ESO Alonso de Cordova 3107, Vitacura, Santiago, Chile, \and Leibniz-Institut f\"ur Astrophysik Potsdam (AIP), An der Sternwarte 16, 14482 Potsdam, Germany, } 

   \date{Received XXX}

% \abstract{}{}{}{}{} 
% 5 {} token are mandatory
 
  \abstract
   {}
   {We present the Lensed Lyman-Alpha MUSE Arcs Sample (LLAMAS) selected from MUSE and HST observations of 17 lensing clusters. The sample consists of 603 continuum-faint (-23<$M_{\rm UV}$<-14) lensed Lyman-$\alpha$ emitters (producing 959 images) with  secure spectroscopic redshifts between 2.9 and 6.7. Combining the power of cluster magnification with 3D spectroscopic observations, we are able to reveal the resolved morphological properties of 268 Lyman-$\alpha$ emitters.}
   {We use a forward modelling approach to model both Lyman-$\alpha$ and rest-frame UV continuum emission profiles in the source plane and measure spatial extent, ellipticity and spatial offsets between UV and Lyman-$\alpha$ emission. }
   {We find a significant correlation between UV continuum and Lyman-$\alpha$ spatial extent. Our characterization of the Lyman-$\alpha$ haloes indicates that the halo size is linked to the physical properties of the host galaxy (SFR, Lyman-$\alpha$ equivalent width and Lyman-$\alpha$ line FWHM). We find that 48\% \  of Lyman-$\alpha$ haloes are best-fitted by an elliptical emission distribution with a median axis ratio of $q=0.48$. 
   We observe that 60\% of galaxies detected both in UV and Lyman-$\alpha$ emission show a significant spatial offset ($\Delta_{Ly\alpha-UV}$). We measure a median offset of $\Delta_{Ly\alpha-UV}= 0.58 \pm 0.14$ kpc for the entire sample. By comparing the spatial offset values with the size of the UV component, we  show that 40\% \ of the offsets could be due to star-forming sub-structures in the UV component, while the larger offsets (60\%) are more likely due to larger distance processes such as scattering effects inside the circumgalactic medium or emission from faint satellites or merging galaxies. Comparisons with a zoom-in radiative hydrodynamics simulation of a typical Lyman-$\alpha$ emitting galaxy show a very good agreement with LLAMAS galaxies and indicate that bright star-formation clumps and satellite galaxies could produce a similar spatial offsets distribution. }
   {}

   \keywords{Galaxies: evolution, galaxies: high-redshift, gravitational lensing: strong}

   \maketitle
%
%-------------------------------------------------------------------

\section{Introduction}

The existence of bright Lyman-$\alpha$ radiation emitted by galaxies was originally predicted by \citet{Partridge1967} and progressively became a prominent target in searches for high redshift galaxies. Then, the detection of extended Lyman-$\alpha$ emission around high redshift galaxies was predicted by \citet{Haiman2000}. The Lyman-$\alpha$ emission is now also used as a neutral hydrogen gas tracer in the circumgalactic medium (CGM) given its resonant nature, which produce extended halos surrounding galaxies up to 30 kpc (\citealt{Matsuda2012, Momose2014}), with the exception of quasars.

The origin of extended Lyman-$\alpha$ haloes surrounding galaxies is still unknown with two main hypotheses being considered: scattering of Lyman-$\alpha$ photons, produced mostly within star-forming regions, through the interstellar (hereafter ISM) and circum-galactic medium or in-situ photoionisation emission or collisional emission in the CGM (\citealt{Mitchell2021}). Lyman-$\alpha$ haloes therefore represent a powerful probe of the hydrogen gas within the CGM, tracing both spatial extent and velocity structure of the gas surrounding galaxies and thus investigating the galaxy formation processes and the reionisation epoch at $z=6$.
These resonant scattering events increase the path length of Lyman-$\alpha$ photons and consequently the observed Lyman-$\alpha$ emission is potentially influenced along this path by a large number of physical parameters (column density, temperature, dust content, kinematics, covering fractions and clumpiness), which modify both the spectral profile of the line and its spatial distribution \citep{Ouchi2020}.\\

Observing and characterizing Lyman-$\alpha$ haloes is crucial to understand the nature of the CGM at low and high redshifts. The LARS collaboration (for Lyman-$\alpha$ Reference Sample, \citealt{Hayes2013, Hayes2014,Ostlin2014}) characterised the Lyman-$\alpha$ emission in low redshift star-forming galaxies ($z<0.5$), demonstrating the presence of a complex structure in the outer parts of the disks. At high redshift ($z>2$), conducting similar studies is harder because of limitations in sensitivity and spatial resolution. Narrow-band imaging observations noted that $z>2$ galaxies appear more extended in Lyman-$\alpha$ than in the rest-frame UV continuum (\citealt{Moller1993,Fynbo2001}).  \citet{Hayashino2004} later detected the first extended halo by stacking 74 Lyman-$\alpha$ emitters (hereafter LAEs) at $z=3.1$ using the Subaru Telescope, while \citet{Rauch2008} also observed 27 faint LAEs between $z=2.6$ and $3.8$ in very deep long slit exposures with ESO-VLT, finding that the majority of their LAEs had spatial profiles larger than the UV sources. More recently, \citet{Steidel2011} stacked the images of 92 bright Lyman Break Galaxies (LBGs) at $z=2.3-3$, and demonstrated that Lyman-$\alpha$ emission is detected out to 10" from the UV continuum emission at a surface brightness level of $\sim 10^{-19} \rm \ erg \ s^{-1} \ cm^{-2} \ arcsec^{-2}$. This stacking approach was adopted by other groups (\citealt{Matsuda2012, Feldmeier2013, Momose2014, Xue2017, Wu2020}) and confirmed the presence of extended Lyman-$\alpha$ haloes with typical exponential scale lengths of $\sim 5-10$ kpc. However, a study of  the possible diversity among  individual Lyman-$\alpha$ haloes remained out of reach in these observations due to the lack of sensitivity.\\

The arrival on sky of the ESO-VLT instrument MUSE (the Multi Unit Spectroscopic Explorer, \citealt{Bacon2010}), an integral-field spectrograph with an unrivalled sensitivity, has substantially increased the number of observed LAEs at high redshift ($z>3$) (\citealt{Bacon2015}). The sample was extended with the MUSE {\it Hubble} Ultra Deep Field (UDF, \citealt{Bacon2015, Leclercq2017}) and the MUSE Wide Survey (\citealt{Lutz2016}). \citet{Leclercq2017} reported the detection of individual extended Lyman-$\alpha$ emission around 80\% of the LAEs detected, confirming the presence of a significant amount of hydrogen gas in the CGM. They presented a systematic morphological study of 145 of these Lyman-$\alpha$ haloes in the UDF, showing that the majority of the Lyman-$\alpha$ flux comes from the halo surrounding each galaxy, whose properties seem to be related to the UV properties of the galaxies. In parallel, the Integral Field Unit (IFU) instrument KCWI (Keck Cosmic Web Imager, \citealt{Morrissey2018}) at the Keck Observatory, recently started to confirm similar results in $2<z<3$ LAEs  (\citealt{Erb2018, Chen2021}). \\

Scenarios about the origin of the extended Lyman alpha emission include cooling radiation from cold accretion, outflows, emission from satellites galaxies and resonant scattering of photons produced in the ISM and the CGM (\citealt{Laursen2007, Steidel2010, Zheng2011}). Spatially integrated Lyman-$\alpha$ emission is almost always redshifted relative to the systemic velocity, indicating the presence of galactic outflows in both observational and theoretical works (\citealt{Heckman2001, Verhamme2006, Scannapieco2017,Song2020}). Cosmological simulations predict also the presence of inflowing filamentary streams of dense gas (\citealt{Keres2005, Dekel2006, Mitchell2021}) which could produce an overall blue-shifted Lyman-$\alpha$ line (\citealt{Dijkstra2006, Verhamme2006, Mitchell2021, Garel2021}).
Galaxies with a relatively low Lyman-$\alpha$ optical depth typically exhibit double-peaked profiles, with a dominant red peak, with the peak separation strongly dependent on the kinematic of the gas and neutral hydrogen column density (\citealt{Verhamme2006, Henry2015, Gronke2016,Verhamme2017}). In addition, emission from satellite galaxies and in-situ emission probably contribute to the spatial extent and the clumpy morphology of the Lyman-$\alpha$ haloes (\citealt{Mas-Ribas2016, Mas-Ribas2017,Mitchell2021} in simulations).

Although the spatial resolution of the MUSE deep fields in \citet{Leclercq2017}   data did not allow them to distinguish between the aforementioned scenarios, by comparing these data with a zoom-in cosmological simulation, \citet{Mitchell2021} demonstrated that simulated Lyman-$\alpha$ haloes  are likely powered by a combination of scattering of galactic Lyman-$\alpha$ emission, in-situ emission of the (mostly infalling) CGM gas  and Lyman-$\alpha$ emission from small satellite galaxies. In their simulation, \citet{Mitchell2021} showed that each of the scenarios is dominant on a different scale.
Another approach to study the physical processes influencing the Lyman-$\alpha$ emission is to use simple wind models, in which the central Lyman-$\alpha$ source is surrounded by an expanding neutral hydrogen medium associated with dust. Albeit simple in nature, these models can successfully reproduce the majority of observed Lyman-$\alpha$ line profiles (\citealt{Ahn2004, Schaerer2008, Verhamme2008, Schaerer2011, Gronke2015, Yang2016, Gronke2017, Song2020}). \\

One outstanding feature of Lyman-$\alpha$ haloes is the spatial offset between UV continuum and Lyman-$\alpha$ emission peaks repeatedly reported in the litterature (\citealt{Shibuya2014, Hoag2019, Lemaux2020, Ribeiro2020}). Most studies so far used long-slit spectroscopy to measure offsets between UV and Lyman-$\alpha$ emission, but these observations are restricted to one dimension only. Taking advantage of the 3D spectroscopy, IFUs are a more efficient tool to provide a more thorough and detailed picture of the Lyman-$\alpha$ and UV emission in individual galaxies.
The presence of spatial offsets could indicate that Lyman-$\alpha$ photons can preferentially be produced (or scattered) far away from the star forming regions that are responsible for the UV emission. Depending on the range of offsets observed they could preferentially support one of the two main scenarios mentioned previously. For example, an offset smaller than the UV source extent could indicate an off-center star-formation clump emitting a large amount of Lyman-$\alpha$ photons. On the contrary an offset larger than the UV source size will support the scenario of satellite galaxy emission or resonant scattering from escape channels. Measuring these spatial offsets very precisely and comparing them with the UV emission distribution represent a great opportunity to study the preponderance of different scenarios. 
\\

Multiple studies have focused on bright/extreme Lyman-$\alpha$ haloes, such as  \citet{Swinbank2015} who noted a variation of hydrogen column density in the CGM at $z=4.1$. Similar studies were performed by \citealt{Erb2018} (a double peaked Lyman-$\alpha$ line at $z=2.3$), \citealt{Vernet2017} (at $z=3$), \citealt{Matthee2020, Matthee2020_2} (an LBG at $z=6.53$ and one galaxy at $z=6.6$) and \citealt{Herenz2020} (one Lyman-$\alpha$ blob at $z=3.1$). \citet{Leclercq2020} performed a similar study of six bright haloes from the total UDF sample, searching for resolved variations of the Lyman-$\alpha$ line profile across the halo and correlations with host galaxy properties. They showed that the Lyman-$\alpha$ line is in general broader and redder in the extended part of the halo, suggesting that Lyman-$\alpha$ haloes are powered either by scattering processes in an outflowing medium. However the lack of sufficient spatial resolution (typically $\sim 3-4.9 \ \rm kpc \ at \ z=4$) of these studies left the physical interpretation widely open, and some questions are still pending, namely: What are the origins of the Ly$\alpha$ photons? Which physical mechanisms (e.g., outflows, inflows, satellites galaxies) are responsible for the extent of Lyman-$\alpha$ haloes and the spectral shapes of the Lyman-$\alpha$ line profiles?\\

In order to improve our understanding of the properties of the CGM and provide robust new constraints on theoretical models, we focus our observations on high-redshift lensed galaxies. %Gravitational lensing boosts 
Gravitational lensing boosts and magnifies the total observed flux of sources, causing them to appear physically larger and (in some cases) creating multiple images of a single object, making them ideal targets for spatially resolved studies. A small but growing number of highly magnified LAEs have already been individually studied 
in order to characterize the CGM gas at $2<z<7$. One of the first is a lensed galaxy at $z=4.9$ presented in \citet{Swinbank2007}, showing extended Lyman-$\alpha$ emission. Following this study, many subsequent %observations 
efforts have also targeted strongly lensed sources (\citealt{Karman2015, Caminha2016, Patricio2016, Vanzella2016, Smit2017, Claeyssens2019, Vanzella2020, Chen2021}). These studies, focusing only on one or two objects, have provided the first evidence of 
variations in the Lyman-$\alpha$ line profile across the halo, revealing %and 
the complex structure of the neutral hydrogen distribution surrounding galaxies (in terms of covering fraction, column density, presence of inflows/outflows). Besides these studies have demonstrated that lensing observations represent a privileged field to study the CGM at high redshift. However, most studies of lensed galaxies only concern too few objects to draw general conclusions. \\

With this in mind, we construct a statistically large sample of lensed Lyman-$\alpha$ emitters named the Lensed Lyman-$\alpha$ MUSE Arc Sample (hereafter LLAMAS). Based on the recent MUSE lensing clusters data release presented in \citet{Richard2021} (hereafter R21), totalling 141 hours were obtained mainly through MUSE Guaranteed Time Observation (GTO) on 18 different fields; we construct a unique sample of 603 lensed LAEs (forming 959 images) at $z=2.9-6.6$. Their strong lensing properties are characterised using well-constrained models of massive galaxy clusters, with magnification values ranging from 1.4 to 40. A partial sample of these LAEs was used previously to study the lens models (e.g. \citealt{Mahler2018,Lagattuta2019}) and the Lyman-$\alpha$ luminosity function in individual / several clusters (\citealt{Bina2016, GDLV}). The LLAMA sample makes use of the unique combination of strong lensing, deep MUSE IFU and HST high resolution images to study the spatial and spectral properties of the Lyman-$\alpha$ haloes in the entire sample. In this first publication of the series, we present the general properties of this sample and focus on the morphological parameters, with emphasis on the spatial extent of the Lyman-$\alpha$ emission and spatial offsets between UV continuum and Lyman-$\alpha$ emission. In order to physically interpret these results and disentangle between the many Lyman-$\alpha$ halo production scenarios, we compare our results with a zoom-in radiation-hydrodynamical  simulation. The paper is organized as follows: we describe our data and the sample selection in Sect.~\ref{sec:data}. Section~\ref{sec:Analysis} presents our procedure for image construction, spectral extraction and fitting, and modeling of the UV and Lyman-$\alpha$ spatial distribution in the source plane. We describe the results in  Sect.~\ref{sec:results}. We discuss these results and compare with the zoom-in simulation in Sect.~\ref{sec:discussion}. Finally we present our summary and conclusions in  Sect.~\ref{sec:conclusion}. 
All distances are physical and we use AB magnitudes. We adopt a $\Lambda$ cold dark matter cosmology with $\Omega_{\Lambda}=0.7$, $\Omega_{m}=0.3$ and $H_0=70 \ {\rm km}\ {\rm s}^{-1} \ {\rm Mpc}^{-1}$.

%--------------------------------------------------------------------
\section{Lyman-$\alpha$ sample}
\label{sec:data}
The data and  redshift catalogs presented in this study have all been processed following the method described in R21. Among the 17 galaxy clusters catalogs used for this work, 12 were presented in R21 and 5 are presented for the first time (MACS0451, MACS0520 are completely new and  A2390, A2667 (\citealt{GDLV}) and AS1063 (\citealt{Mercurio2021}) had been the subject of previous studies). This section will shortly describe the method presented in R21 and the LAEs selections based on the global catalogs.
\subsection{MUSE data}
\label{sec:MUSE_data}
The MUSE data reduction procedure details are given in R21, largely following the prescription described in \citet{Weilbacher} with some specific improvements for crowded fields. 
Of the 17 clusters we explore here, five (MACS0451, MACS0520, A2390, A2667 and AS1063) were analysed after the publication of R21, but the final data products and galaxy catalogs were constructed following exactly the same procedure. The final output is given as a FITS datacube with 2 extensions containing the flux and the associated variance over a regular 3D grid at a spatial pixel scale of $0.2$" and a wavelength step of 1.25 \AA\ between 4750 and 9350 \AA. The final seeing, defined as the FWHM of the point spread function at $7000$ \AA, varies from  0.52" to 0.79" among the fields.
%The 17 clusters have  
Every cluster in our sample has a redshift between $0.2$ and $0.6$ and are all known to be massive strong lenses. The integration times vary between 2 and 14 hours per field, using a combination of standard and adaptive-optics (for observations done after 2014) modes.
The fields of view are centered on the core regions of 
each cluster, in order to maximise the number of strongly lensed LAEs. The MUSE field of view is 1 $\times$ 1 arcmin$^2$; for five clusters, multiple contiguous MUSE pointings were mosaicked to cover the complete multiple image area of the clusters (between 2 and 4 pointings, see Table \ref{tab:clusters_table}).

\subsection{HST data}
\label{sec:HST_data}
To complement the MUSE data we use the available high-resolution ACS/WFC and WFC3-IR images in the optical / near-infrared covering the MUSE observations. 
Six clusters in the sample are included in either the CLASH \citep{Postman2012} or Frontier Fields \citep{Lotz2017} surveys, and therefore have deep HST observations taken in 12 and 6 filters, respectively. For six additional clusters, HST observations were obtained as part of MACS survey (PI: Ebeling) as well as follow-up HST programs (PIs: Bradac, Egami, Carton). 

\subsection{Input redshift catalogs}
\label{sec:catalog}
Based on these observations, a complete spectroscopic catalogue was constructed for each cluster. The complete procedure is described in R21 and the main steps are:
    \begin{itemize}
        \item Production of an input photometric catalogue of continuum sources that overlap with the MUSE field of view. The detection image (produced by combining the HST images into an inverse- variance-weighted detection image) is given as input to the {\sc SExtractor} software \citep{Sextractor}.  The photometry performed by SExtractor in each band is merged together into a final catalogue of HST sources (hereafter PRIOR sources).
        \item Independent of the HST catalogue, a line-detected sources catalogue is produced directly from the MUSE datacubes, performed by running the {\sc muselet} software which is part of the {\sc MPDAF} python package (\citealt{mpdaf}) hereafter MUSELET sources.
        \item A spectrum of each source (both PRIOR and MUSELET) is extracted from the MUSE datacube based on weighted images (created for each source by taking the flux distribution over the segmentation maps produced as part of the detection process) to optimise the signal-to-noise of the detections. A local background spectrum  around each source was estimated and subtracted (to remove large-scale contamination from bright sources, such as stars and cluster members, as well as potential systematics in the background level remaining from the data reduction) from the weighted spectrum to compute an optimized spectrum for source identification and redshift measurement.
        \item All MUSELET and PRIOR sources down to a limiting signal-to-noise ratio (S/N) in the MUSE continuum (averaged over the full MUSE wavelength range) were inspected individually and their redshift were assessed. For each source we determine the redshift value and confidence (between 1 for insecure redshifts and 3 to most secure redshifts, the details of redshift classification can be found in R21 Sect 3.5), as well as any association between PRIOR and MUSELET sources (i.e. when the same source is detected in both HST and MUSE data). \item The resulting catalogs are then tested with the corresponding {\sc lenstool} \citep{Lenstool} mass model  of the cluster to associate multiple images together and predict potential new multiple images. 
        \item The final catalogs are composed of 4020 secure redshifts with $0<z<6.7$ and 634 unique LAEs with redshift confidence $>1$ (more details on the Lyman-$\alpha$ line identification can be found in R21 Sect. 3.5)\footnote{https://cral-perso.univ-lyon1.fr/labo/perso/johan.richard/MUSE\_data\_release/}. The redshifts of the LAEs are measured based on the Lyman-$\alpha$ emission line or the presence of a strong Lyman break or from nebular lines when detected. 
    \end{itemize}
\begin{table*}
    \centering
    \begin{tabular}{l|rrclllll}
Cluster & R.A. & Dec. & $z_{\rm cl}$ & MUSE depth & N pointings & N LAEs & \textbf{$V_{\rm eff, \ \mu>1.5}$} & \textbf{$V_{\rm eff, \ \mu>5.4}$}   \\

 & (J2000) & (J2000) & & [hours]&  & & [Mpc$^3$] & [Mpc$^3$] \\
 \hline
Abell 2744  &  00:14:20.702  & $-$30:24:00.63  &  0.308 & 3.5 - 7 & 4 & 142 (\textbf{121}) & {5080.4} & {84.4}\\
Abell 370  &  02:39:53.122 & $-$01:34:56.14  &  0.375 & 1.5 - 8.5 & 4 & 98 (\textbf{42}) & {3566.5} & {216.8}\\
MACS\,J0257.6$-$2209  &  02:57:41.070 & $-$22:09:17.70  &  0.322 & 8 & 1 & 48 (\textbf{25}) & {711.7} &{76.9}\\
MACS\,J0329.6$-$0211  &  03:29:41.568 & $-$02:11:46.41 &  0.450 & 2.5 & 1 & 8 (\textbf{17}) & {1155.8} & {63.9}\\
MACS\,J0416.1$-$2403 N  &  04:16:09.144 & $-$24:04:02.95  &  0.397 & 17 & 1 & 71 (\textbf{46}) & {1330.0} & {45.3}\\
MACS\,J0416.1$-$2403 S  &  04:16:09.144 & $-$24:04:02.95  &  0.397 & 11-15 & 1 & 56 (\textbf{34}) & {1330.0} & {45.3}\\
MACS\,J0451.9$+$0006 &  04:51:54.647 & $+$00:06:18.21  &  0.430 & 8 & 1 & 45 (\textbf{21}) & {863.5} & {51.3}\\
MACS\,J0520.7$-$1328 & 05:20:42.046 & $-$13:28:47.58 & 0.336 & 8 & 1 & 33 (\textbf{19}) & {696.6} & {101.4} \\
1E 0657$-$56 (Bullet)  &  06:58:38.126 & $-$55:57:25.87  &  0.296 & 2 & 1 & 14 (\textbf{11}) & {898.3} & {69.0}\\
MACS\,J0940.9+0744  &  09:40:53.698 & $+$07:44:25.31  &  0.335 & 8 & 1 & 58 (\textbf{49}) & {2310.1} & {20.2}\\
MACS\,J1206.2$-$0847  &  12:06:12.149 & $-$08:48:03.37  &  0.438 & 4-9 & 3 & 82 (\textbf{50}) & {2791.2} & {133.8}\\
RX\,J1347.5$-$1145  &  13:47:30.617 & $-$11:45:09.51 &  0.451 & 2-3 & 4 & 124 (\textbf{72}) & {2929.4} & {96.8}\\
SMACS\,J2031.8$-$4036 &  20:31:53.256 & $-$40:37:30.79  &  0.331 & 10 & 1 & 44 (\textbf{21}) & {1329.9} &{55.4}\\
SMACS\,J2131.1$-$4019  &  21:31:04.831 & $-$40:19:20.92  &  0.442 & 7 & 1 & 30 (\textbf{16}) & {586.0} & {100.3}\\
Abell 2390 & 21:53:36.823 & $+$17:41:43.59 & 0.228 & 2 & 1 & 14 (\textbf{8}) & {759.3} & {98.1}\\
MACS\,J2214.9$-$1359  &  22:14:57.292 & $-$14:00:12.91  &  0.502 & 7 & 1 & 33 (\textbf{17}) & {699.6} & {51.6}\\
Abell S1063 & 22:48:43.975 & $-$44:31:51.16 & 0.348 & 3.9& 2 & 35 (\textbf{20}) & {1839.9} & {146.3}\\
Abell 2667 & 23:52:28.400 & $-$26:05:08.00 & 0.233 & 2 & 1 & 24 (\textbf{14}) & {820.0} & {99.0}\\
\hline
Total & & & & & 30 & 959 (\textbf{603}) & {29534.8} & {1556.7} \\

 \end{tabular}
 \medskip\par
    \caption{Summary of the 17 galaxy clusters. The {seventh} column shows the number of Lyman-$\alpha$ images detected in each field with a redshift confidence of 2 or 3 and a Lyman-$\alpha$ line with S/N>3. The boldface values are the number of unique LAEs detected in each cluster with high redshift confidence. The last columns indicates the effective volume surveyed ($V_{\rm eff}$) at $2.9 < z < 6.7$ with MUSE for Lyman-$\alpha$ emitters, {for a magnification of the brightest image higher than $1.5$ and $5.4$, respectively.}}
    \label{tab:clusters_table}
 \end{table*}
 
\subsection{Mass models}
\label{sec:mass_models}
In order to study the intrinsic galaxy properties of the LAEs in the source plane, we used the parametric models of the cluster mass distribution presented in R21, generated using the public {\sc lenstool} software \citep{Lenstool}. 
{\sc lenstool} allows to generate parametric models of each cluster's total mass distribution, using numerous multiple images identified in the catalogs as constraints.
The final model's parameters and constraints are presented in Appendix B of R21. Each cluster mass model is optimized with between 7 and 100 multiples systems of images with secure spectroscopic redshifts. 
The precision of the lens models, which corresponds to the typical spatial uncertainty in reproducing the strongly-lensed images, is typically from 0.5" to 0.9".
One crucial value for the study of lensed background source morphologies is the lensing amplification and shear. We use the value from R21 as a first estimate, based on the central location of each image in the catalogs, and refine it in the Section 3.5.
As {\sc lenstool} uses a Monte Carlo Markov Chain (MCMC) to sample the posterior probability distribution of the model, statistical errors were estimated for each parameter of the models.
A second important measurement derived from the lens model is the equivalent source-plane area covered by the MUSE observations. The intrinsic survey volume for lensing studies differs from the image-plane area due to strong lensing effects. The effective survey volume is reduced by the same amount as the magnification factor. This value varies depending on the cluster (due to the different mass distribution and MUSE coverage)  and redshifts of the sources. At $z=4$, which is the median redshift of all the LAEs detected in this sample, the total source-plane area covered is about 1.8 arcmin$^2$. The relative contribution of each cluster to the full survey covolume is provided in the last columns of Table \ref{tab:clusters_table}.

\subsection{LAE selection}
\label{sec:LAE_selection}
The catalogs presented in R21 include all the spectroscopic redshifts measured in each field. In order to construct a sample of LAEs, we selected all the sources with a secure redshift (confidence 2 or 3 based on multiple emission lines, a clear asymetric Lyman-$\alpha$ emission line, a clear Lyman break or a lensing confirmation of the high-redshift nature of the image) between 2.9 and 6.7 (1031 Lyman-$\alpha$ images selected on 1269 detected).
We flagged all galaxies for which no Lyman-$\alpha$ emission is detected (e.g. with an Lyman-$\alpha$ line integrated S/N$>$3). For these sources, we searched for extended Lyman-$\alpha$ emission in multiple NB images produced around the predicted location of the Lyman-$\alpha$ line (based on the galaxies' systemic redshift) with different velocity windows. We rejected 20 galaxies with no significant emission features around the galaxy location (S/N$<$3).
After a visual inspection we rejected all images detected in close vicinity of a bright cluster galaxy (BCG) or bright cluster members. In such cases the Lyman-$\alpha$ line is too  contaminated by the foreground galaxy continuum and the HST emission is not sufficiently isolated to be correctly spatially fitted. In fact, all the images rejected are either central poorly magnified images or small counter-images from a multiple system. The rejection of these images does not affect the final sample results since either the images would have a too low S/N to be spatially fitted (see Section.~\ref{sec:lenstool_selection}) or they are parts of a multiple system of which the other images are kept in the sample.
The final LLAMAS catalogue is composed of 959 Lyman-$\alpha$ images from 603 unique objects. Among these sources,  341 have at least one image with an HST detection. The number of LAEs detected in each cluster is presented in Table \ref{tab:clusters_table} together with observation information on each field.

\subsection{Global properties}
\label{sec:global_properties}
\begin{figure*}
    \includegraphics[width=18cm]{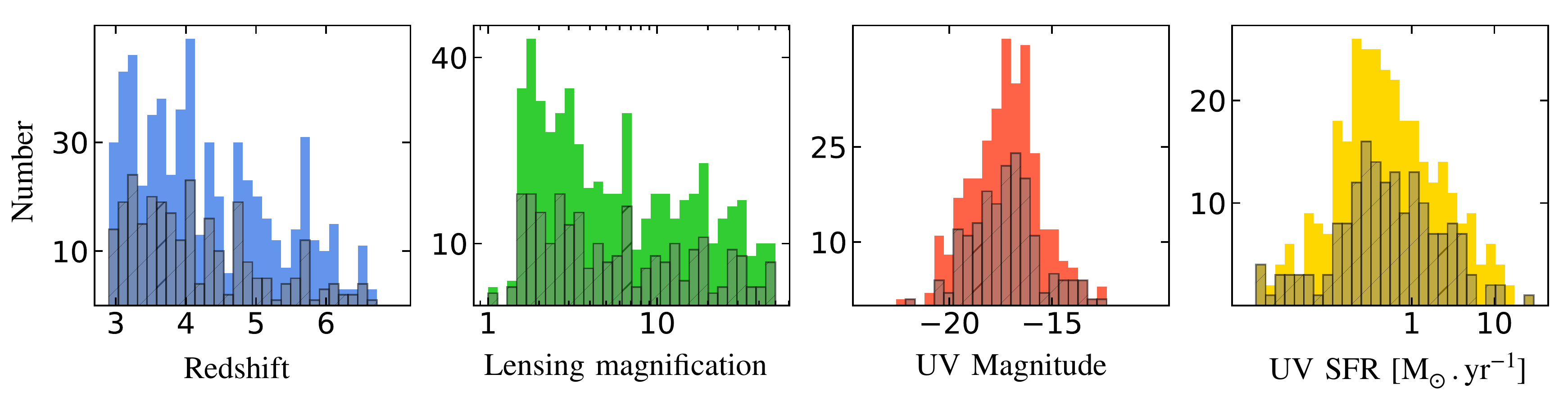}
    \caption{Global properties of the galaxies from the LLAMAS. From left to right: distribution of redshift, lensing magnification, UV magnitude and SFR (both corrected from lensing magnification). The grey area in each panel represents the sources selected for the spatial fitting (see Section.~\ref{sec:lenstool_selection}). When more than one image is detected for the same source, the values of the most magnified images are presented in panels 3 and 4. The magnifications presented in this figure are the total magnification for each source, i.e the sum of the global magnifications of each multiple image of the same galaxy.}
    \label{fig:hist_global}
\end{figure*}

Figure \ref{fig:hist_global} shows the global properties of the LLAMAS: redshift, lensing magnification, UV magnitude and UV star formation rate (SFR) distributions, the last two corrected for lensing magnification. The grey area of each panel represents sources selected for the spatial fitting (see Section~\ref{sec:lenstool_selection}). 
The median redshift of the complete sample is $z=4$ with  40 galaxies at $z>6$. The median magnification value is $\mu=5.4$, with a range from 1 to 40. The magnification values presented in Figure\ref{fig:hist_global} are the values computed in R21, based on the central position of each image.
The UV absolute magnitude at rest-frame 1500 \AA\ (M$_{1500}$) and the UV spectral slope ($\beta$) are estimated by adjusting a single power law to the HST broad-band photometry available in each cluster (see R21) redward of Lyman-$\alpha$ in the UV. The UV SFR is derived from M$_{1500}$ using the \citet{Kennicutt98} relation. Among the sample, 39$\%$ of the objects are pure MUSELET detections (without any UV detection in HST). For these objects, no UV magnitude or SFR can be computed.
Among the PRIOR sources, 65$\%$ of the galaxies have ${\rm SFR} <1 \  M_\odot\,{\rm yr}^{-1}$, and the median value is ${\rm SFR} =0.55 \, M_\odot {\rm yr}^{-1}$. 

%-----------------------------------------------------------------------------------
%-----------------------------------------------------------------------------------
%-----------------------------------------------------------------------------------

\section{Analysis}
\label{sec:Analysis}
\subsection{Narrow-band image construction and global spectral properties}
\label{sec:NB_images}

To characterize the Lyman-$\alpha$ emission of each LAE we first construct a 5"$\times$5" Lyman-$\alpha$ narrow-band (hereafter NB) image of each object from the MUSE datacube. When a source produces multiple images, we study each image independently. With the intention of maximising the S/N in the NB images  and recovering most of the Lyman-$\alpha$ emission of each source, we applied an iterative process consisting in three steps: spectral fitting, NB image construction, and spectral extraction repeated three times.  \\

First we fitted the Lyman-$\alpha$ line with the asymmetric Gaussian function (\citealt{Shibuya2014_2}) given by:
\begin{equation}
\label{eqn:asym}
    f(\lambda) = A \ {\rm exp}(\frac{-(\lambda -\lambda_{\rm peak})^2}{2(a(\lambda- \lambda_{\rm peak})+d)^2})
\end{equation}
with $A$ the amplitude of the line, $\lambda_{\rm peak}$ the peak of the line, $a$ the asymetry parameter and $d$ the typical width of the line.
Four parameters were optimised: peak position ($\lambda_{\rm peak}$) of the line, FWHM, flux $F$, and asymmetry $a$ of the profile. 
The prior on the flux corresponds to the integrated flux of the last produced spectrum on the spectral range $[\lambda_{\rm peak} -6.25: \lambda_{\rm peak} +6.25]$ \AA.
The prior on $\lambda_{\rm peak}$ is based on the catalogue redshift of each source. When two spectral peaks were detected in the Lyman-$\alpha$ line, the spatial fit only took the red peak into account. We applied a uniform prior of FWHM and asymmetry with mean values of $7$ \AA\ and $0.20$, respectively. We applied this fit on a spectral window around the Lyman-$\alpha$ emission peak, in which each pixel has a minimum signal to noise ratio of 2.5.
Each spectrum and its associated variance were fitted using the python package {\sc emcee} \citep{emcee}. We performed the fit with 8 walkers and 10000 steps. We used the median values of the resulting posterior probability distributions for all the model parameters as best-fit parameters. Errors on the parameters were estimated using the 16th and 84th percentiles.  In the first loop of the process, this spectral fit was applied on the continuum optimised, sky-subtracted spectrum produced in the data release of R21. \\

Secondly we constructed a Lyman-$\alpha$ narrow-band image for each Lyman-$\alpha$ image. The central wavelength is based on the results of an asymmetric Gaussian function fit of the Lyman-$\alpha$ line. The continuum level was measured on the left and right side of the Lyman-$\alpha$ line, over a 28 \AA \ width band and subtracted from the collapsed Lyman-$\alpha$ line image. The NB bandwidth $[ \lambda_{\rm left}: \lambda_{\rm right}]$, with $\lambda_{\rm left}$ and $\lambda_{\rm right}$ being the left and right wavelengths of the band where the spectral layers were summed, was optimised (from $\lambda_{\rm right}- \lambda_{\rm left}=$ 2.5 to 20 \AA) to maximise the S/N as measured in a 0.7" radius circular aperture (typical MUSE PSF size) centred on each image (on the catalogue source position). The start and end position of the spectral band for continuum level estimation were  chosen to be close to the Lyman-$\alpha$ line while not containing any Lyman-$\alpha$ emission ($[\lambda_{\rm left} -34 \ \AA \ : \lambda_{\rm left} -6.25 \ \AA ]$ on the left side and $[\lambda_{\rm right} +6.25 \ \AA \ : \lambda_{\rm right} +34 \ \AA ]$) on the right side. Following this procedure, we obtained NB images with spectral bandwidths ranging from 3.75 to 17.5 \AA \ depending on the galaxy (which represent 3 to 14 MUSE spectral pixels, respectively). The mean spectral width is 6.9 \AA \ which represents 345 $\rm km .s^{-1}$ at $z=4$. \\

Third, the new Lyman-$\alpha$ image was obtained, a non-weighted spectrum was re-extracted from the MUSE datacube based on this new NB image. The purpose of this new extraction is to get a Lyman-$\alpha$ optimised spectrum containing the most of the Lyman-$\alpha$ emission from the galaxy and its halo. We spatially smoothed the best NB image with a Gaussian filter of $\rm FWHM_{\rm smooth}= 0.4"$. The total spectra is the sum of the spectra of each MUSE pixel with a flux in the NB image higher that the typical value of the dispersion measured in the image. To avoid external contamination, a mask was created manually for each object to isolate them from possible cluster galaxies or star residuals in the NB images. Finally the sky subtraction was performed using the same sky spectrum used to create the first spectrum. \\

The same process was performed two more times, each time based on the last NB image and extracted spectra.
If an object presents a Lyman-$\alpha$ emission too low for the spectral extraction (i.e. no pixels with a smoothed SB level $\rm > 6.25 \ 10^{-19} \ erg.s^{-1}.cm^{-2}.arcsec^{-2}$ in any of the tested NB images, the object was rejected from the sample (in total 12 images for 5 objects rejected).

\subsection{Forward modelling}
\label{sec:forward_modelling}
\begin{figure}
    \includegraphics[width=9cm]{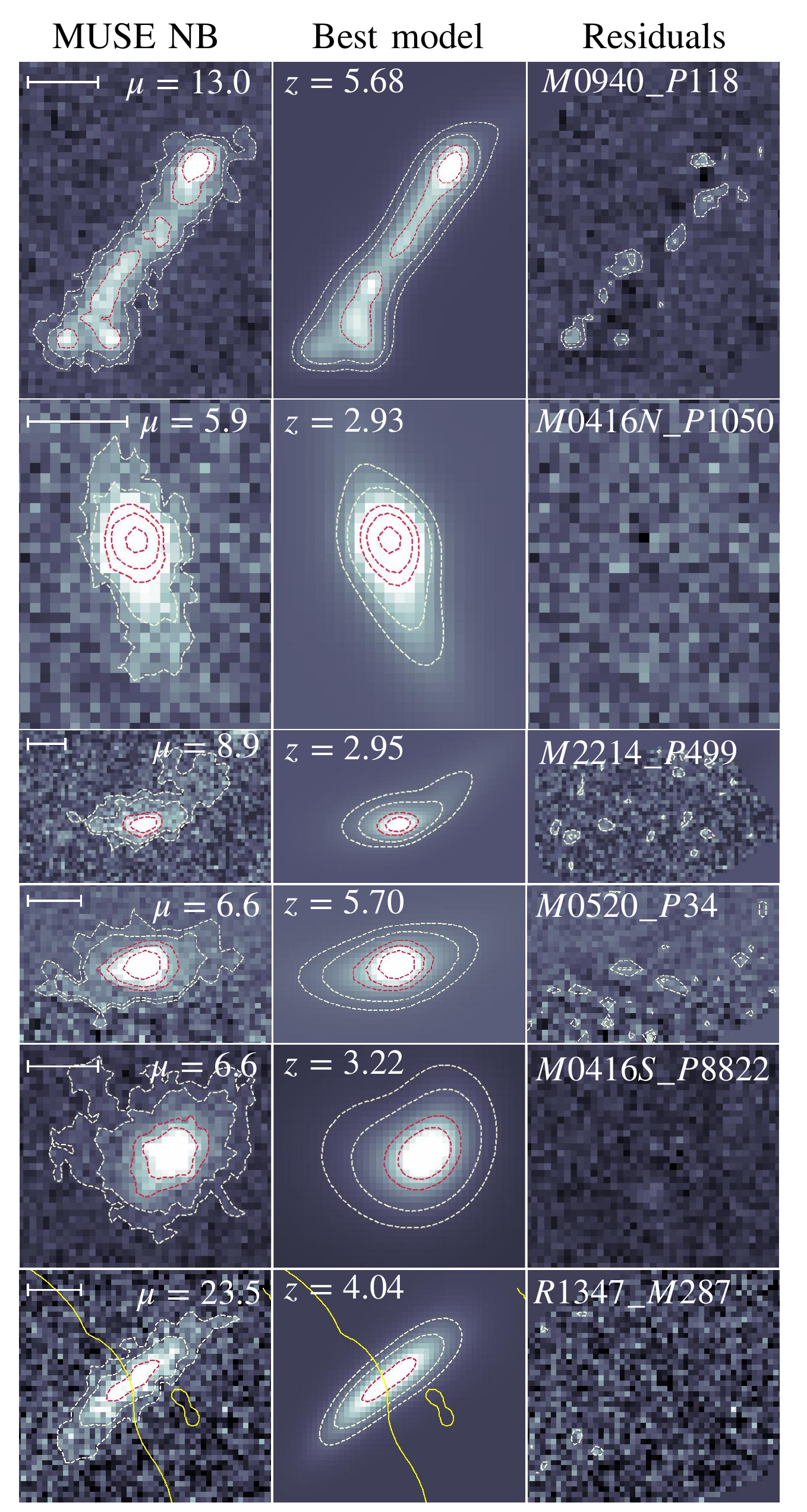}
    \caption{Examples of LLAMAS galaxies and spatial best models. In each row, we show, from left to right, the MUSE NB image, the best model and residuals. The contours present smoothed surface brightness levels at $12.5$ and $25.0 \times 10^{-19} \rm \ erg\ s^{-1} \ cm^{-2} \ \rm arcsec^{-2}$ in light yellow and at $62.5$, $100$ and $200 \times 10^{-19} \rm \ erg \ s^{-1} \ cm^{-2} \ \rm arcsec^{-2}$ in red. The scale in the left panels are all at 2 arcseconds. We indicated from left to right the magnification value ($\mu$), the redshift and ID of the source. In the last column, the areas without pixels indicates the edges of the area located in the region where the fit is applied, when any area without pixels is visible, it means that the region used is larger than the image presented in the figure. In the last row, the yellow line represent the critical line.}
    \label{fig:exemple_fits}
\end{figure}

To study intrinsic morphological properties of the LAEs, we modeled the Lyman-$\alpha$ and UV continuum emission in the source plane making use of the cleanlens function from the latest version of  {\sc lenstool}. The method used in this study is the forward modelling approach based on parametric source models, including both lensing and instrumental effects. It consists in generating parametric source models (based on user assumptions on the emission profile) in the source plane, lensing it by the best cluster model, convolving with the PSF and re-gridding it to the spatial sampling of the observation, to compare with the observation. In this study we used only Sersic profiles and specifically exponential (Sersic index $n=1$) as used in \citet{Lutz2016} and \citet{Leclercq2017}. The free parameters are spatial location ($x$ and $y$) of the center, exponential scale radius ($\rm a$), ellipticity ($\epsilon=(a-b)/(a+b)$ with $b$ the minor axis of the distribution), position angle ($\theta$), magnitude ($\rm m$) and Sersic index ($\rm n$, fixed to 1 in case of exponential profiles). The best-fit parameters were found by minimizing the residuals between input observed images and simulated image-plane observations. One or several Sersic components could be used to reproduce one object. The multiple images of the same object could be fitted together or separately. 
In this study we decided to fit each multiple image separately to avoid multiple images models uncertainties effect on the source reconstruction. We applied this fit both on Lyman-$\alpha$ emission from NB images and UV continuum emission from HST images for objects with an HST detection in the catalogue.
To isolate each object from other features presented in the NB images, we constructed manually a contour around each image, extended enough to cover all significant flux pixels (see Figure ~\ref{fig:exemple_fits}) and a large (at least 10 MUSE pixels around the image) empty area around it. Only pixels inside this region are considered  for the $\chi^2$ calculation. A weight image is associated for each object. For MUSE observations the weight values were estimated as $\rm 1/Var[p,q]$ in each pixel (p,q) from the NB variance image associated. For the HST images, the standard deviation $\sigma$ was measured in an empty region close to the object, and the values of all pixels of the weight image were fixed to $1/\sigma^2$. The model includes a contribution from a local background ($sky$) estimated from the median flux measured in a large empty region close to each source.
The $\chi^2$ estimate in {\sc lenstool} is then:
\begin{eqnarray}
\label{chi2_SM}
\chi^{2} = \sum\limits_{p,q} {\rm (I_{p,q} - (M_{p,q}+{\it sky}))}^2  \rm W_{p,q} \   ,
\end{eqnarray}
with $\rm I_{p,q}$, $\rm M_{p,q}$ and $\rm W_{p,q}$  respectively the value in pixel [p,q] of the observed, model and weight images.
\subsection{PSF estimation}
\label{sec:PSF}
Because we aim at modelling morphological properties of the LAEs, we have to obtain a very good knowledge of the point spread function (PSF) in both HST and MUSE observations. Since the PSF varies with wavelength across the MUSE spectral range we determined a specific circular Moffat monochromatic PSF for each object with an FWHM estimated following equation 1 in R21. We constructed MUSE PSF following the same procedure described in \citet{Bacon2017}. The approach to determine PSF FWHM variations with wavelength for each cluster is described in R21. 
For the HST images, we modeled the PSF in each filter used for spatial fitting (F555W, F606W, F814W, F110W and F125W). We used at least 5 non-saturated, bright and isolated stars detected in each cluster (in all filters). The HST images of all these stars were combined to create a 51 $\times$ 51 pixels average image of the PSF centred on the brightest pixel used as PSF by {\sc lenstool}.

\subsection{Validation and selection with {\sc lenstool}}
\label{sec:lenstool_selection}

\begin{figure*}
    \includegraphics[width=18cm]{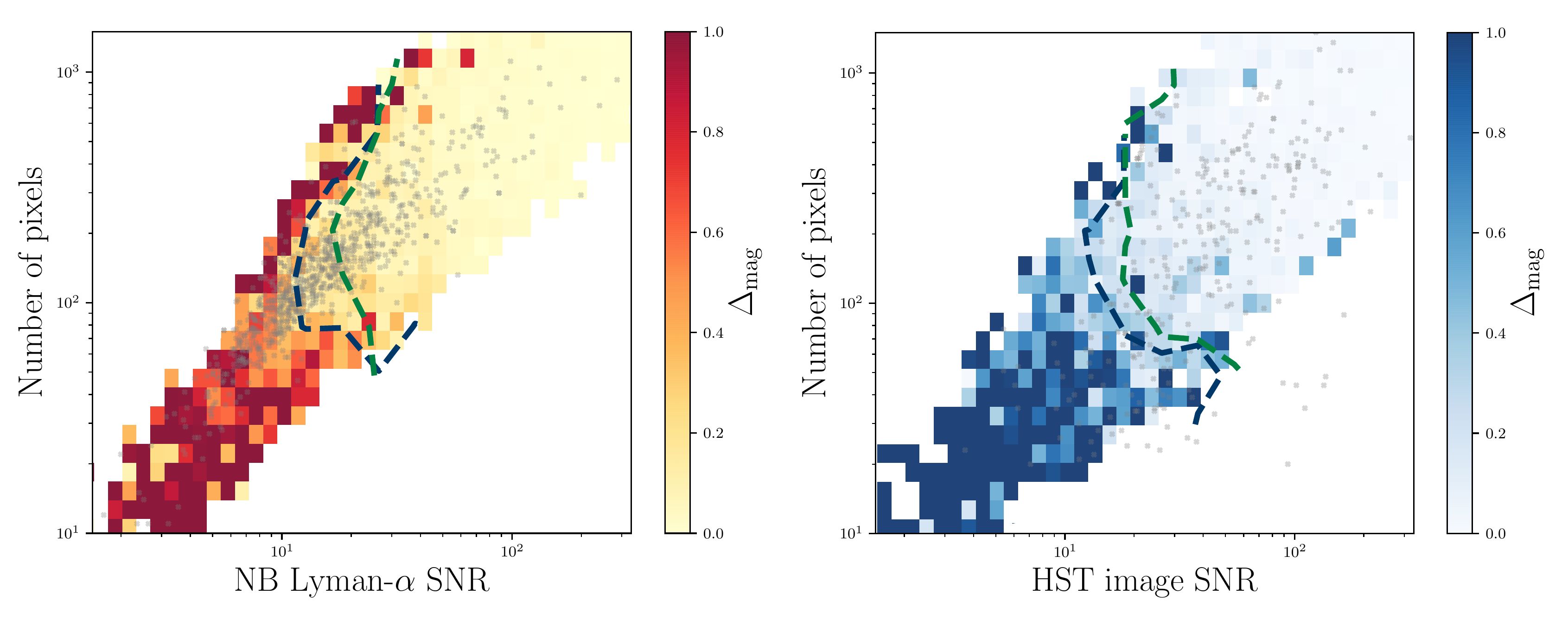}
    \caption{Distribution of residual magnitudes between simulated source and best-fit parameters for MUSE Lyman-$\alpha$ NB images in the left panel and HST images in the right panel (see section 3.4). The grey points represent the LLAMAS galaxies. All the objects in the blue dashed contour were selected for spatial fitting with {\sc lenstool} based on the magnitude difference between input and best-fit parameters ($\Delta_{\rm mag}<0.3$). The green dashed contour represents where the spatial fit will recover both the flux distribution and morphological information on the sources as estimated by the fraction of non-overlap between the simulated and fitted sources (see Sect.~\ref{sec:lenstool_selection}).}
    \label{fig:select_llamas_flux}
\end{figure*}

To estimate the robustness of the source-plane modeling as a function of the S/N and extent of images we fitted a range of simulated Sersic profiles with the same method applied on real LAEs (for both MUSE and HST images). We generated more than 4000 simulated sources Sersic profiles with randomly selected parameters (position, scale radius, ellipticity, positional angle and magnitude). Each source was lensed by a real cluster model (4 randomly selected clusters from R21) and convolved with the MUSE PSF. We added random realizations of the noise based on different noise measurements from the Lyman-$\alpha$ NB images. 
Once a simulated image was created, we detected the multiple images using the Python/ {\sc photutils} package \citep{Bradley2016}. If multiple images of the source were detected, only one image was fitted (chosen randomly). After applying the forward modelling approach described before, the best-fit parameters were compared with initial source parameters. Figure~\ref{fig:select_llamas_flux} shows how the difference between input and best-fit magnitudes varies as a function of S/N and area of each simulated image. We consider that a difference $<0.3$ in magnitude is enough to get a good representation of the flux distribution in the source plane (with a relative error lower than 5\% and 10\% on the scale length and ellipticity parameter respectively). A region of the plot stands out visually, we defined a contour at the level of $\Delta _{\rm mag} = 0.3$ that we used after as a selection function for the source-plane spatial study. The S/N and the number of pixels contributing to the spatial fit were measured on the optimized NB image of each object with exactly the same detection process applied on simulated sources. The total distribution of the complete sample of LAEs is represented with grey points in Figure~\ref{fig:select_llamas_flux}.
Finally we obtained 475 MUSE Lyman-$\alpha$ images and 271 objects selected for source-plane emission spatial characterisation. Among them, 142 objects have enough resolved HST data to be characterised both in UV and Lyman-$\alpha$ emission (which represent 206 images).

The green line on Figure~\ref{fig:select_llamas_flux}, represents the criterion used to select sources for which morphology best-fit parameters are reliably recovered. To determine if the morphology of a source is correctly recovered by the fit procedure we compared in the source plane the input and best-fit models. Both input and best-fit profiles are elliptical profiles, to measure the difference between the 2 ellipses, we determined the proportion of non-recovered morphology with respect to the area of the input ellipse by measuring: $ R_{\rm eps} = \frac{(e_{\rm 1} + e_{\rm 2} - 2\ (e_{\rm 1} \cap e_{\rm 2}))}{e_{\rm 1}}$ with $e_{\rm 1}$ the area of the input ellipse, $e_{\rm 2}$ the area of the best-fit ellipse and $e_{\rm 1} \cap e_{\rm 2}$ the common area. The two ellipses are defined by their scale radius, ellipticity and positional angle and the position. We consider the morphological properties measured well enough when this $R_{\rm eps} < 0.3$, e.g. the error on morphology of the source concerns less than 30\% of the total source area. The signal-to-noise ratio used for this selection is integrated on the complete image. Images with a good signal-to-noise ratio (between 10 and 35) but a very large number of pixels (between 200 and 1000) can not be well fitted as the signal-to-noise ratio of each individual pixel is relatively low.

\subsection{Modelling the morphology in the source plane}
\label{sec:source_plane_fit}
The large majority of previous studies on the SB spatial distribution of LAEs (\citealt{Steidel2011,Lutz2016,Leclercq2017}) mainly characterised the morphological properties of the Lyman-$\alpha$ nebulae through 1 or 2 components circular exponential models. \citet{Leclercq2017} decomposed the spatial fitting of the Lyman-$\alpha$ emission in 2 steps: first they fitted the UV emission on HST images using a single circular exponential profile. Then they fitted the Lyman-$\alpha$ emission on MUSE NB images using a combination of 2 circular exponential profiles, at the same location, one with the scale radius fixed to the UV one. This modelling reproduced well the objects of these studies, with good residuals. But in our study, thanks to the lensing magnification, we observe LAEs with an improved spatial resolution. After applying this two circular exponential components model on all the Lyman-$\alpha$ images selected for spatial fitting, it was obvious that this model was not suitable for a large fraction of the LLAMAS galaxies. Indeed a lot of lensed galaxies from our sample present either a two-components Lyman-$\alpha$ distribution or a strongly asymmetric fainter Lyman-$\alpha$ emission surrounding a bright core emission. 
With the intention of obtaining the best source-plane reproduction for the UV and Lyman-$\alpha$ emission distribution we chose to apply between 11 and 7 different models, respectively presented in Table~\ref{tab:fits_description}.

 First if the object is detected in HST with enough S/N to be selected for the spatial fitting (see Figure~\ref{fig:select_llamas_flux}), we performed 2 fits on the UV image: the first is a circular exponential profile (model M1) and the second an elliptical exponential profile (model M2). We applied the fit on the F555W, F606W, F814W, F110W or F125W filter (depending of target redshift, quality of the detection in terms of S/N of the images and available HST data in each cluster) whichever was the closest to the 1500-2500 \AA \ restframe. Then we fitted 9 models on the MUSE Lyman-$\alpha$ emission: 
   \begin{itemize}
       \item \textbf{M3}. The classical two-component circular exponential profile based on UV modelling (5 free parameters)
       \item \textbf{M4}. The same two-component exponential profile with elliptical profiles based on the elliptical UV modelling (7 free parameters)
       \item \textbf{M5}. The M1 profile  with the possibility to adjust the UV-like component scale radius (6 free parameters)
       \item \textbf{M6}. The M3 profile with the possibility to adjust the UV-like component scale radius, ellipticity and position angle (10 free parameters)
       \item \textbf{M7}. Two component circular exponential profiles allowed to get centroids at 2 different locations (8 free parameters)
       \item \textbf{M8}. Two component elliptical exponential profiles allowed to get centroids at 2 different locations (12 free parameters)
       \item \textbf{M9}. A single-component circular exponential profile (4 free parameters)
       \item \textbf{M10}. A single-component elliptical exponential profile (6 free parameters)
       \item \textbf{M11}.  A single-component Sersic elliptical profile (7 free parameters)
   \end{itemize}
 The complete description of free and fixed parameters used in each model is presented in Table~\ref{tab:fits_description}.

To disentangle which modelling is the more adapted for each object we took into account the number of constraints, the number of free parameters and the final $\chi^2$, using a Bayesian Information Criterion (BIC) defined as: $\rm BIC = -2ln(\mathcal{L})  -k \ ln (N) $ with $\mathcal{L}$ the likelihood of the best-fit, k the number of free parameters and N the number of constraints. We kept as the best modelling the fit with the minimum BIC.
   
Thanks to these different modellings, we obtained for each LLAMAS galaxy the best source-plane spatial model of both UV (when it is detected) and Lyman-$\alpha$ emission. Table \ref{tab:fits_table} shows how sources are distributed among the best-fit models for each cluster and for the complete sample. We divided the different Lyman-$\alpha$ emission fits in 3 categories: two components fixed at the same location, two components free to vary, and the single component models. In the end, 67\% of the LLAMAS galaxies are well described with a  two-components fixed model (but the two-components circular exponential model is chosen only for 15 \%). 12 \% of the galaxies are best described by two free components, which correspond to LAEs presenting either 2 Lyman-$\alpha$ emission spatial peaks or a very asymmetric Lyman-$\alpha$ distribution.

We looked for trends between the best-fit model and the source S/N or lensing magnification effects on the models distribution. We measured median S/N of the objects in each category of models and found that the two more complex fits (models 7 and 8 with 2 free components) have a  median S/N of twice as high as the median S/N of the other models (S/N=25). Finally we found that objects modelled with one of the two free components fits have a median lensing magnification of 6.2 versus 3.7 for the two fixed components models and 4.3 for the one component models. Both S/N and magnification seem to impact the models distribution between the different fits. The fact that high S/N and high magnification objects tend to prefer a more complex fit to reproduce intrinsic emission distribution suggests that LAEs have a more complex structure than a circular Lyman-$\alpha$ halo surrounding a circular ISM Lyman-$\alpha$ emission component. 
   
\begin{figure}
    \includegraphics[width=8cm]{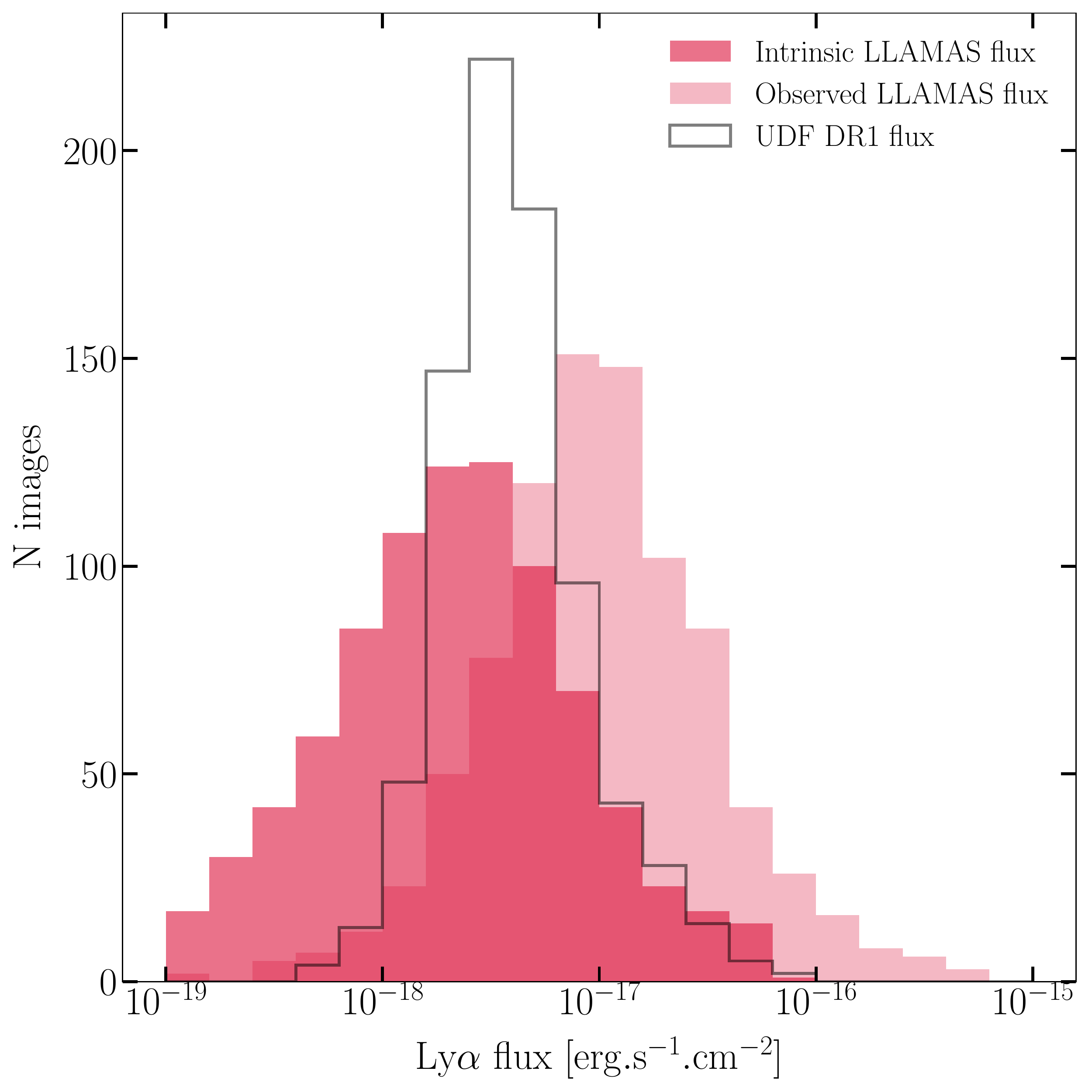}
    \caption{Lyman-$\alpha$ line flux distribution of the LLAMA galaxies and UDF Lyman-$\alpha$ haloes presented in \citet{Leclercq2017}. The light pink histogram shows the distribution of the observed flux of Lyman-$\alpha$ images in the LLAMA sample. The dark pink histogram represents the intrinsic flux distribution of the LLAMA galaxies, obtained by dividing the observed flux by the lensing magnification. This histogram shows only the images selected for spatial fitting (see Figure~\ref{fig:select_llamas_flux}). The typical uncertainty at 1 $\sigma$ for LLAMAS flux measurement is about $1.2 \times 10^{-18} \ \rm erg.s^{-1}.cm^{-2}$.  The grey histogram shows the observed Lyman-$\alpha$ flux of the Lyman-$\alpha$ haloes presented in \citet{Leclercq2017}.} 
    \label{fig:fluxdistrib}
\end{figure}

Based on the spectral fit performed on the Lyman-$\alpha$ line (see Section.~\ref{sec:NB_images}) we obtain a measure of the total Lyman-$\alpha$ emission flux for each image. Figure \ref{fig:fluxdistrib} shows the distribution of observed and intrinsic Lyman-$\alpha$ flux of the LLAMAS galaxies selected for spatial fitting. The grey histogram shows the distribution of the total Lyman-$\alpha$ line flux of the 145 LAEs presented in \citet{Leclercq2017}. Thanks to the lensing magnification, our sample allows the characterisation of fainter LAEs than non-lensed studies.
To estimate the more realistic intrinsic flux, we use the best-fit of each object to update the lensing magnification value to account for the morphology of each source. The magnification was estimated by measuring the ratio between Lyman-$\alpha$ emission in the image plane (measured in the best-fit image plane results) and the source-plane emission (measured in the best-fit source-plane model). These new magnification values better represent the total amplification of the Lyman-$\alpha$ emission compared to the previous values measured at a specific UV location, especially for the most strongly magnified sources. We find that for magnification under 10, the two magnification values are very similar; for higher magnification values, the new measurement is on average 2 to 10 times lower than the first estimate, which was an expected outcome. Indeed, for highly magnified images, the magnification varies across the image and thus a value measured at one position in the image does not reflect the average magnification of this image.

\begin{table*}
    \centering
    \begin{tabular}{l|c|l|l}
Model description & N components &  Fixed parameters & Free parameters\\
 \hline
 
 \textbf{M1}. One circular exponential & 1 & $\epsilon=0$ ; n=1 & x; y; a; m \\
 
 \textbf{M2}. One elliptical exponential & 1 &  n=1 & x; y; a; m; $\epsilon$; $\theta$ \\
 
 \textbf{M3}. Two circular exponential & 2 & $\epsilon_1=\epsilon_2 =0$; $n_1=n_2=1$; & $x_1=x_2$; $ y_1=y_2$; \\
 fixed components & & $\rm a_{\rm 1}= a_{\rm UV}$ & $\rm a_{\rm 2}$ ;$m_1$; $m_2$ \\
  
 \textbf{M4}. Two elliptical exponential  & 2 & $\epsilon_1= \epsilon_{\rm UV}$; $\theta_1=\theta_{\rm UV}$ & $x_1=x_2$; $y_1=y_2$; $\rm a_{\rm 2}$ \\
 fixed components & &$\rm a_1= a_{\rm UV}$; $n_1=n_2=1$; & $m_1$; $m_2$; $\epsilon_2$; $\theta_2$ \\
 
 \textbf{M5}. Two circular exponential & 2 & $\epsilon_1=\epsilon_2 =0$; $n_1=n_2=1$& $x_1=x_2$; $ y_1=y_2$; \\
 fixed components & &  & $\rm a_{\rm 1}= a_{\rm UV} \pm 0.1 "$; $\rm a_{\rm 2}$ ;$m_1$; $m_2$; \\

 \textbf{M6}. Two elliptical exponential & 2 & $n_1=n_2=1$& $x_1=x_2$; $ y_1=y_2$; \\
 fixed components & &  & $\rm a_{\rm 1}= a_{\rm UV} \pm 0.1 "$; $\rm a_{\rm 2}$ ;$m_1$; $m_2$; \\
  & & &$\epsilon_1$; $\epsilon_2$; $\theta_1$; $\theta_2$\\  
  
\textbf{M7}. Two free circular & 2 & $n_1=n_2=1$; $\epsilon_1=\epsilon_2=0$ & $x_1$; $x_2$; $y_1$; $y_2$; $a_1$; $a_2$; $m_1$; $m_2$\\
components &  &  & \\

\textbf{M8}. Two free elliptical & 2 & $n_1=n_2=1$ & $x_1$; $x_2$; $y_1$; $y_2$; $a_1$; $a_2$; $m_1$; $m_2$\\
components &  &  & $\epsilon_1$; $\epsilon_2$; $\theta_1$: $\theta_2$\\

\textbf{M9}. One circular exponential & 1 & $n_1=1$; $\epsilon_1=0$  & $x_1$; $y_1$; $a_1$; $m_1$\\
component &  &  & \\

\textbf{M10}. One elliptical exponential & 1 & $n_1=1$  & $x_1$; $y_1$; $a_1$; $m_1$, $\epsilon_1$, $\theta_1$\\
component &  &  & \\

\textbf{M11}. One Sersic circular exponential & 1 &  & $x_1$; $y_1$; $a_1$; $m_1$; $n_1$\\
component &  &  & \\

 \end{tabular}
 \medskip\par
    \caption{Description of free and fixed parameters used for the different spatial models applied. $x$ and $y$ are the RA and DEC positions, $m$ is the magnitude, $a$ is the scale radius, $n$ is the Sersic index, $\epsilon$ is the ellipticity and $\theta$ is the positional angle. The models M1 and M2 are applied only on the UV images and models from M3 to M11 on Lyman-$\alpha$ images.}
    \label{tab:fits_description}
 \end{table*}

\begin{table*}
    \centering
    \begin{tabular}{l|r|rc|lll}
Cluster & N images &  HST with MUSE & only MUSE & 2 fixed comp& 2 free comp & 1 comp \\
 & \textbf{(N obj)} & & & & &  \\
 \hline
A2744  & 63 \textbf{(49)} & 43 &  20 & 46 & 4 & 13\\
A370 &39 \textbf{(15)} &16 &23 &23 &6 &11  \\
MACS0257  & 27 \textbf{(12)}  & 14 &  13 & 14 & 1 & 11\\
MACS0329 & 7 \textbf{(4)} &3 &4 &6 &1 &0\\
MACS0416N &33 \textbf{(22)} &27 &6 & 25 & 6 & 2 \\
MACS0416S &30 \textbf{(14)} &27 & 3 & 24 & 2 &3 \\
MACS0451 &16 \textbf{(10)} &2 &15 &14 &0&3  \\
MACS0520 &16 \textbf{(10)} &1 &15 &10 &3 &3  \\
BULLET &5 \textbf{(5)} &1 &4 &5 &0 &0 \\
MACS0940  & 31 \textbf{(25)} & 10 & 21 & 20 & 6 & 4\\
MACS1206 & 41 \textbf{(25)} &11 &30 & 24 & 6 & 11  \\
RXJ1347  & 60 \textbf{(32)} & 16 & 44  & 44 & 6 & 10 \\
SMACS2031 & 37 \textbf{(15)}& 11 & 26 & 24 & 7 &6 \\
SMACS2131  & 15  \textbf{(8)} & 4 &  11 &  11 & 4  & 0\\
A2390 & 5 \textbf{(3)} & 5 & 0 & 3 & 1 & 1\\
MACS2214  & 20  \textbf{(8)} & 6 &  14 & 13 & 3 &4\\
AS1063 & 15 \textbf{(7)} & 15 & 0 & 14 & 1 & 0 \\
A2667 & 7 \textbf{(4)} & 1 & 8 & 5 & 1 & 1 \\
TOTAL & 469 \textbf{(268)} & 45\% & 55\% & 67\% & 12\% & 21\%\\

 \end{tabular}
 \medskip\par
    \caption{Summary of the spatial best-fits distribution in each cluster. The first column shows the total number of Lyman-$\alpha$ images selected for spatial fitting and the boldface numbers the number of unique objects. The second and third columns shows the number of images with both UV and Lyman-$\alpha$ detection and only a Lyman-$\alpha$ detection respectively. The three last columns show the distribution in the 3 main categories of Lyman-$\alpha$ emission models. The last \textbf{row} shows the repartition for the global sample.}
    \label{tab:fits_table}
 \end{table*}

%--------------------------------------------------------------------
\section{Results}

In this section we present results on the Lyman-$\alpha$ nebulae morphology: spatial extent and axis ratio of the Lyman-$\alpha$ haloes,  spatial offsets between UV continuum and Lyman-$\alpha$ emissions distributions.
\label{sec:results}

\subsection{Extended emission properties}
\label{sec:extent}

Once we obtain a good model fit for each source, we need a common measurement to compare the spatial extent between the different individual objects. We use half-light ($\rm r_{\rm 50}$) and 90\%-light radii ($\rm r_{\rm 90}$) to characterise both the UV and Lyman-$\alpha$ emissions. For the total sample, 38\% of the LAEs have an  HST counterpart bright enough to be spatially modelled with {\sc lenstool}.
We estimate $\rm r_{\rm 50}$ and $\rm r_{90}$ on a source-plane image produced from the parametric model (minimizing the BIC) of each source. Depending on the type of models, we measure the two radii from elliptical or circular rings. For objects with a circular best model (models M3, M5 and M9) we take  $\epsilon = 0$, for elliptical models M4, M6 and M11 we take the axis ratio value of the brightest component, and for the models M7, M8 (with 2 components at different locations) or, M10, we use the mean axis ratio measured from the model M10 (one elliptical exponential component).
We randomly produce 200 source-plane images selected from the {\sc lenstool} MCMC samples and the error bars are obtained by measuring $\rm r_{\rm 50}$ and $ \rm r_{\rm 90}$ on  each image. Error bars are estimated from the 68 \% confidence interval on each side of the best model value. We also measure in the same way the $\rm r_{\rm 50}$ and $\rm r_{\rm 90}$ radius on the best-model images of the Lyman-$\alpha$ haloes from \cite{Leclercq2017}. \\

Based on the $r_{\rm 50,Ly\alpha}$ and $r_{\rm 90,Ly\alpha}$, we measure the concentration parameter of the Lyman-$\alpha$ emission, which measures how compact the Lyman-$\alpha$ light profile is. We measure values ranging from $c_{Ly\alpha}={r_{90}}/{r_{50}}=33.3$ to  $c_{Ly\alpha}=1.15$ with a median value of $2.57$. The median value of $c_{Ly\alpha}=2.57$ corresponds to a Sersic index of $n=1.2$ which is close to the exponential profile index value $n=1$. The concentration of Lyman-$\alpha$ emission is only weakly correlated with the Lyman-$\alpha$ extent, the more compact haloes are also the smaller.

We used the 90-light radius to compare the spatial extent of the Lyman-$\alpha$ emission in LLAMAS and UDF galaxies (\citealt{Leclercq2017}), as the half-light radius of UDF galaxies will be overdominated by the bright continuum-like component and will not reflect the extended halo properties.
Figure~\ref{fig:extended2} shows the distribution of the circularised Lyman-$\alpha$ 90\%-radius $ r_{\rm 90, Ly\alpha}$ as a function of UV $ r_{\rm 90, UV}$ (for multiply-imaged systems we present the value of the most extended image for both UV and Lyman-$\alpha$ emission). Among the LLAMAS, 40\% of sources have a Lyman-$\alpha$ halo substantially smaller than the vast majority (i.e. 97\%) of the halos in \citet{Leclercq2017}.
We measured the inverse-variance weighted mean  of the ratio $x_{90}={r_{\rm 90,Ly\alpha}}/{r_{\rm 90,UV}}$ in both UDF and LLAMAS and find that the LLAMAS galaxies present on average a higher value, with $  x_{\rm 90,\mu,LLAMAS}=18.0$, compared to the UDF sample for which we find $x_{\rm 90,\mu,UDF}=10.40$. The value of $x_{\rm 90,\mu,UDF}$ is consistent to the values measured in \citealt{Leclercq2017} between the halo component scale radius and the core UV-like component.  These values show that there is a large diversity of Lyman-$\alpha$ emission concentrations (very diffuse or peaked), but in 97\% of the cases we observe a Lyman-$\alpha$ halo more extended than the UV continuum emission, and 75\% and 47\% are significantly more extended at 1  and 3 $\sigma$ respectively. This confirms that the Lyman-$\alpha$ emission is intrinsically more extended than the UV component as measured in previous studies (\citealt{Steidel2010,Leclercq2017, Lutz2018}). We find smaller values for $x_{50}=r_{\rm 50,Ly\alpha}/r_{50,UV}$ with $x_{\rm 50, \mu, \rm LLAMAS}=12.0$ and $x_{\rm 50, \mu,\rm UDF}=4.6$.

\begin{figure}
   \begin{minipage}{9cm}
    \includegraphics[width=9cm]{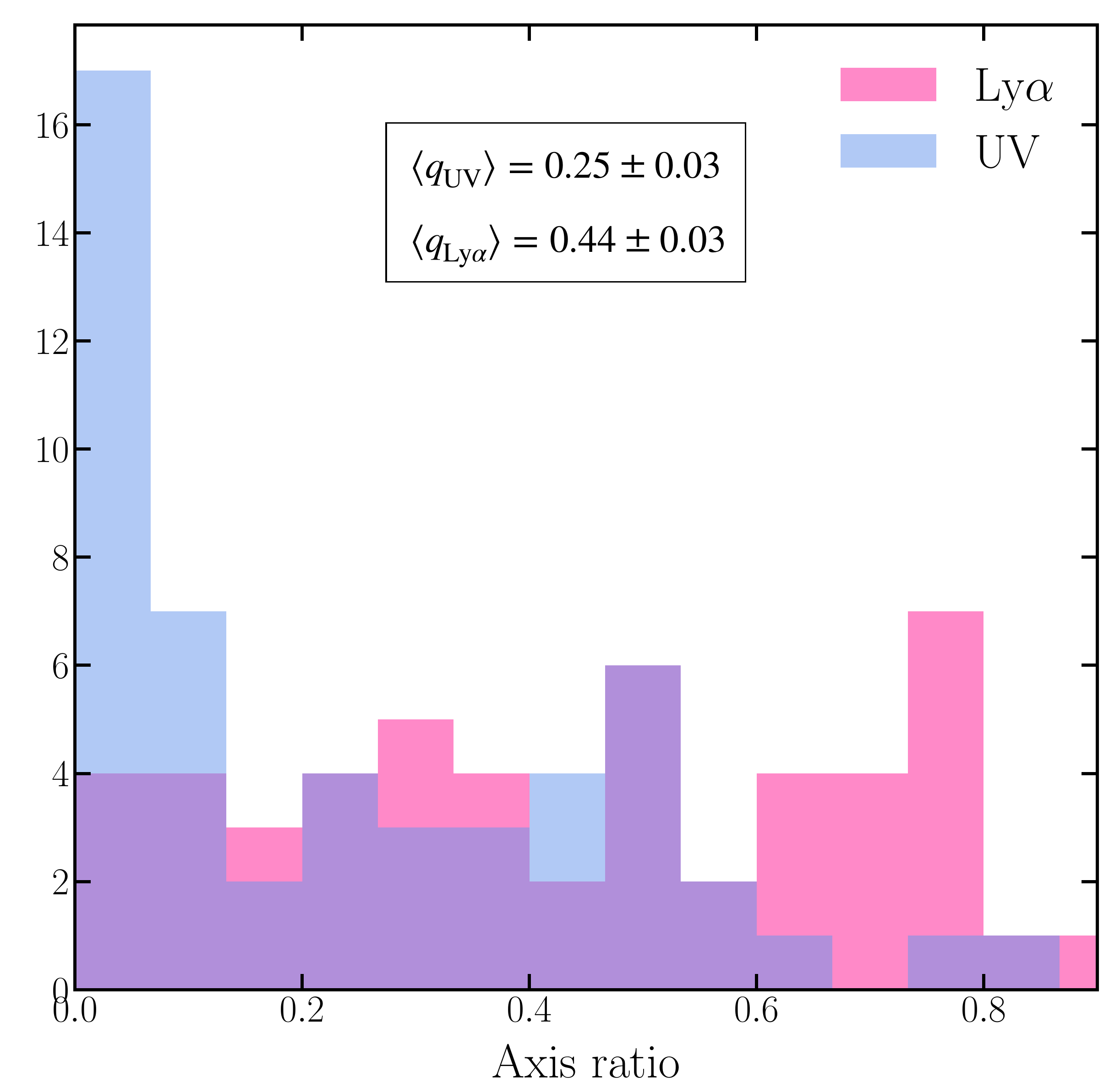}
    \end{minipage}
    \caption{Distribution of axis-ratio ($q=b/a$) of UV (blue) and Lyman-$\alpha$ (pink) emission in the source plane. For multiple systems, we measured the magnification weighted mean of the axis ratio values of only the multiple images located inside the green contour of Figure~\ref{fig:select_llamas_flux}. The mean values of the two subsets are indicated on the plot.  }
    \label{fig:ellipticity}
\end{figure}

From the spatial modelling, we obtain a measure of the axis ratio of the UV and Lyman-$\alpha$ emission distribution (through the axis ratio $q=b/a$ between the major ($a$) and minor ($b$) axes of the best-fit ellipse). This value is interesting to study as it should correlate with the spatial distribution of the neutral hydrogen in the CGM and possibly with the inclination of the galaxy.
We use the best model of each source to determine the degree of ellipticity of the UV and Lyman-$\alpha$ distributions in the source plane. Among the galaxies selected to be spatially modelled in UV emission, 38 \%  prefer a circular best model and 62 \% an elliptical one.  For Lyman-$\alpha$ emission, 52 \% of the haloes are better described by a circular model and 48\% an elliptical one (we consider models M7 and M8 as elliptical models).
The median  UV axis ratio $q$, measured only on elliptical sources, is  $\langle q\rangle\sim0.22$. For multiple systems we measure a magnification weighted mean of the axis ratio values of each image of the system (e.g. the same results are found if we use signal-to-noise ratio weighted mean as the magnification value is strongly correlated with the spatial integrated signal-to-noise ratio). When we consider only images located inside the green contour in the second panel of Figure.~\ref{fig:select_llamas_flux}, we find a median of $\langle q \rangle=0.38$. 
Applying the same procedure to measure the distribution of Lyman-$\alpha$ emission axis ratios, we find that Lyman-$\alpha$ haloes are on average less elongated (median value of $0.22$ for UV and $0.48$ for Lyman-$\alpha$ emission). The distributions of axis ratio values for the  UV and Lyman-$\alpha$ emission distributions are presented in the Figure~\ref{fig:ellipticity}. These results are consistent with the trend presented in \citet{Chen2021} and with the measurements for the high-z simulated LARS galaxies in \citet{Guaita2015} who found similar values of UV and Lyman-$\alpha$ axis ratio  (with axis ratio of their Lyman-$\alpha$ emission between 0.4 and 0.9, with a mean value around 0.7). \citet{Lutz2016} measured also that 75\% of their LAEs came out with a UV axis ratio smaller that 0.5. We use the axis ratio from the best source-plane model and apply the same procedure described previously for the multiple systems. The axis ratio is a useful indicator of the galactic disk inclination and hydrogen distribution morphology for the Lyman-$\alpha$ emission. Haloes with a small axis ratio value indicate that the CGM is structured along a preferred direction around the UV source.\\

We find no significant variation of the axis ratio with redshift. However we found a significant correlation between the spatial extent of the emission in the image plane and the proportion of circular and elliptical best model. We measured the division between circular and elliptical in three equal-size bins based on number of pixels in the detection map. For Lyman-$\alpha$ emission, we measure 13\%, 40\% and 77\% of elliptical best model for respectively $3.3 < {\rm area} < 7$, $7< {\rm area} < 9.6$ and $9.6<{\rm area}<37.5$ with area in arcsec$^2$.  The more a Lyman-$\alpha$ halo is resolved (with a high detection map area which is strongly correlated to the lensing magnification and signal to noise ratio), the more it would prefer an elliptical model. The same effect is observed in S/N as detection map area and S/N are strongly correlated (see Figure~\ref{fig:select_llamas_flux}). These results also indicate that the circular shape measured  on the Lyman-$\alpha$ emission could be partly a limitation due to lower signal to noise ratios and incorrect PSF estimation. \\

\begin{figure*}
    \begin{minipage}{18cm}
    \includegraphics[width=18cm]{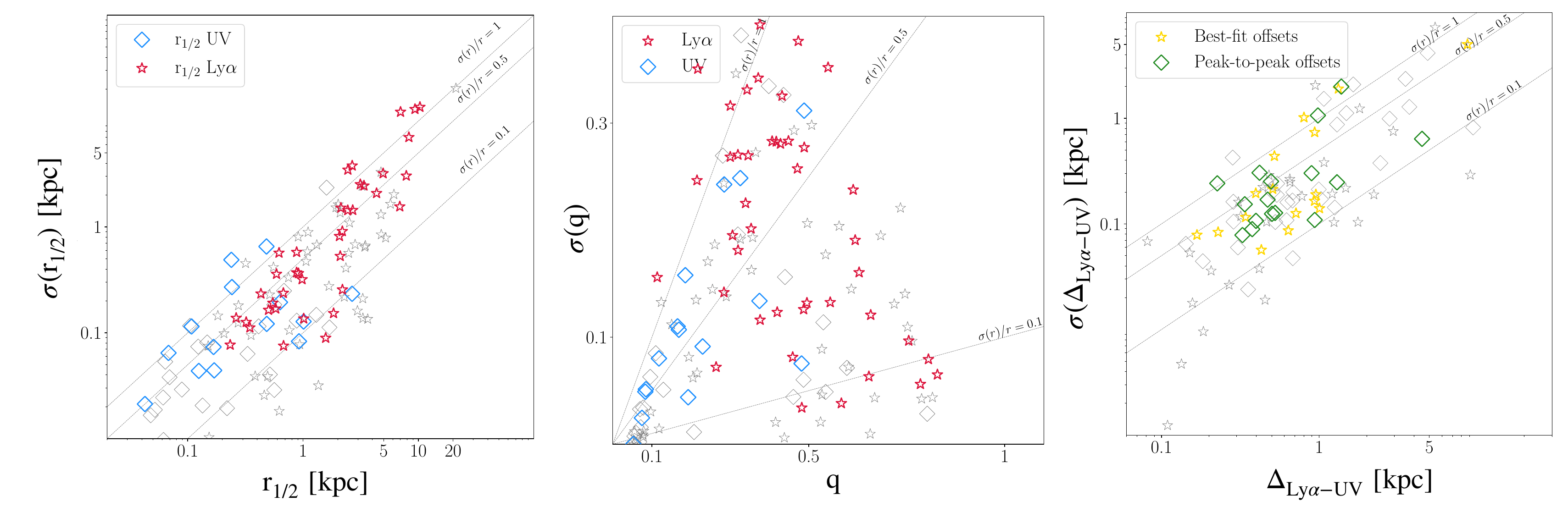}
    \end{minipage}
    \caption{Dispersion measured in each multiple system of the three main parameters measured : UV and Lyman-$\alpha$ extent, axis ratio and spatial offset. We measured in each multiple systems the $\mu$-weighted mean values and $\mu$-weighted dispersion of values among the images. From left to right we represent the dispersion of half-light radius ($r_{\rm 50}$) with respect to $\mu$-weighted mean values, the dispersion of axis ratio ($\rm q$) with respect to $\mu$-weighted mean values and the dispersion of spatial offset ($\Delta_{Ly\alpha-UV}$) with respect to $\mu$-weighted mean values. In each panel the grey stars and squares represent 2 images systems and colored points the 3 and more images systems. In the left and middle panels, the stars represents UV $\rm r_{\rm 50}$ and $\rm q$ respectively, and square Lyman-$\alpha$ values. In the right panel, the stars represent best model centroid offsets and green points represents the "peak to peak" offset values (see Sect.~\ref{sec:offset}). In all panels we plot dashed lines to represent the 10\%, 50\% and 100\% error levels. }
    \label{fig:displensing}
\end{figure*}

Since the lensing models are used to measure all the properties presented here, lensing uncertainties in the mass model used in the lensing reconstruction could strongly impact the results. To estimate the impact of the lensing model on the spatial measurements we study the dispersion of the different measurements between the different images of the 80 multiple systems with at least two images selected for spatial fitting (including 22 systems with 3 or more images with both Lyman-$\alpha$ and UV detections). For each system we measured the magnification-weighted mean ($\langle . \rangle_{\mu}$) and standard deviation ($\sigma_{\mu}$) for each of the following parameters :  $r_{50, Ly\alpha}$, $r_{\rm 50,UV}$, UV ($ q_{\rm UV}$) and Lyman-$\alpha$ axis ratio ($ q_{Ly\alpha}$) and spatial offsets (presented in the following sections). All these measurements are presented in Figure~\ref{fig:displensing}.

For the spatial extent measurements (left panel of the Figure~\ref{fig:displensing}), we found that 19\% of the multiple systems present a small dispersion between the different images (variation smaller than $20\%$ of the mean value in $\langle\rm r_{Ly\alpha}\rangle_\mu$ and in $\langle\rm r_{\rm UV}\rangle_\mu$). For 30\% of the systems, the variation between the different images is moderate (between 20 and 50\% of the mean value) and the remaining 50\% present a large variation between multiple images.  Finally 4 multiples systems present a variation of the Lyman-$\alpha$ extent larger than the mean value. After a visual inspection, we found that these systems are the most magnified galaxies (with a total magnification between 20 and 50). Within these specific systems the variations in amplification and shear, and therefore in spatial resolution, between images are very large, which explains the variations observed in the measurements. In all the results presented in this study, we keep the values measured on the most extended image of each multiple system. We find the same kind of trends for the axis ratio (middle panel of the Figure~\ref{fig:displensing}).

\subsection{Variations in the spatial extent of Lyman-$\alpha$ emission}
\begin{figure}
    \begin{minipage}{9cm}
    \includegraphics[width=8cm]{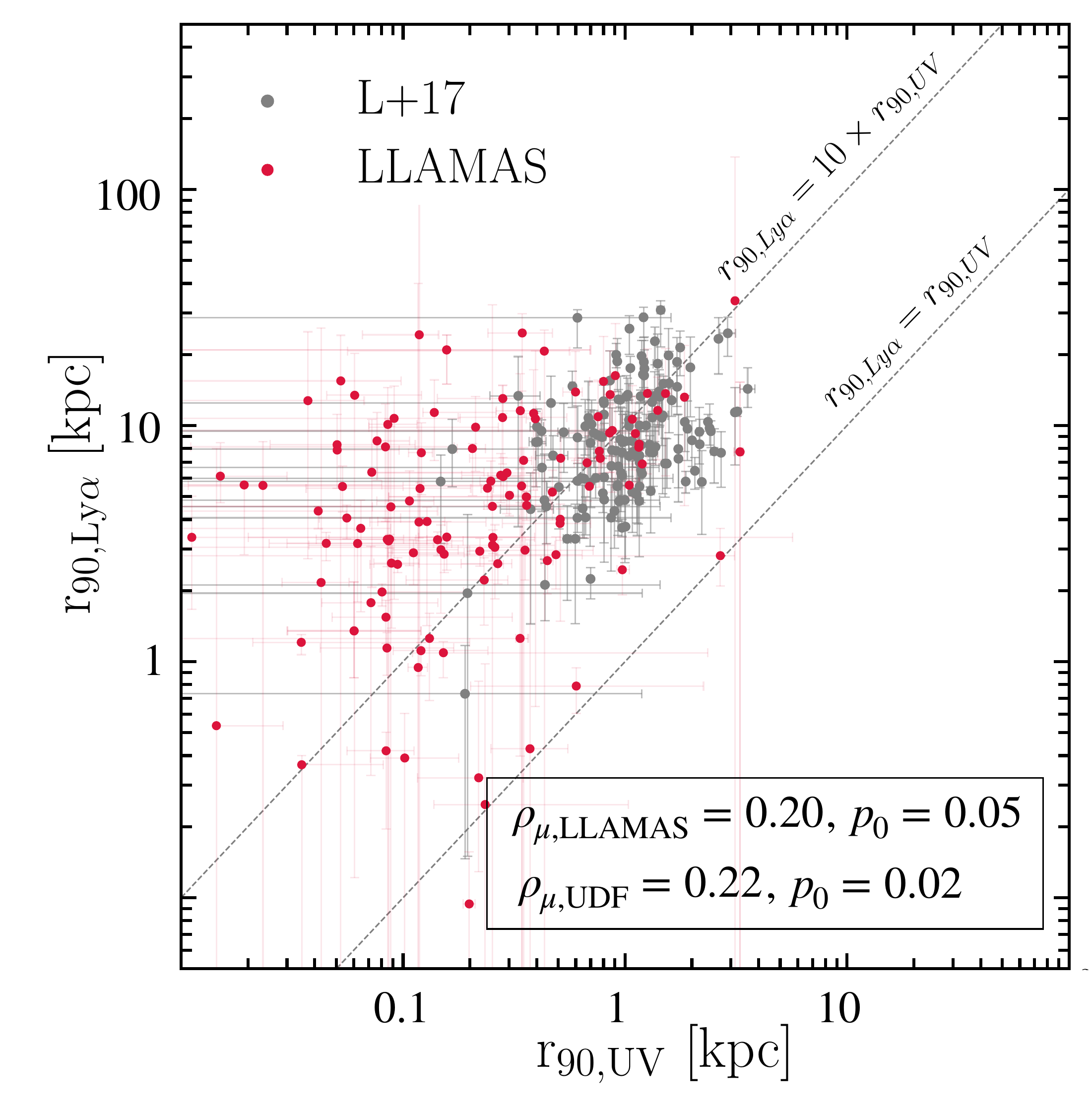}
    \end{minipage}
    \caption{ Comparison of Lyman-$\alpha$ and UV circularised 90$\%$-light radii. The figure shows the Lyman-$\alpha$ 90$\%$-light radius. LLAMAS and UDF \citep{Leclercq2017} galaxies are represented in red and grey, respectively. The UDF and LLAMAS values are both correlated (Spearman weighted  p-value of  $p_0=0.02$ and $p_0=0.05$ for UDF and LLAMAS, respectively). Both datasets are roughly centered around $\rm r_{Ly\alpha}= 10 \ r_{\rm UV}$.}
    \label{fig:extended2}
\end{figure}

In the Figure~\ref{fig:extended2}, we observe a correlation between Lyman-$\alpha$ and UV continuum $\rm r_{\rm 90}$ for both UDF and LLAMAS datasets.
The error-bars weighted Pearson coefficients ($\rho_{UDF}=0.22$ and $\rho_{LLAMAS}=0.20$) suggest that these correlations are weak but significant, with p-values of ($p_{\rm 0,UDF}=0.02$ and $p_{\rm 0,LLAMAS}=0.05$). The LLAMAS values seem more scattered ($\sigma_{\rm UDF}=7.6 \ {\rm kpc} <\sigma_{\rm LLAMAS}=22$  kpc) and the correlation more marginal. This higher spread could be due to the larger uncertainties of the LLAMAS measurements only or also to the fact that lensing studies provide access to a new population of weaker and smaller LAEs than those characterised by the previous studies.
We notice a strong effect of the detection isophotal area (number of pixels in the detection map of each Lyman-$\alpha$ image) on this correlation. We divided the sample in three equal-size bins of low, medium and high value of spatial extent and measured the weighted Pearson p-value and coefficient in each bin. We find that the less extended images (between 3.3 and 6.1 arcsec$^2$) present no correlations between Lyman-$\alpha$ and UV $\rm r_{90}$ ($p_0>0.2$ and $\rho<0.1$). On the contrary the two bins of medium (between 6.1 and 10.2 arcsec$^2$) and highly (between 10.2 and 44.2 arcsec$^2$) extended images show a significant correlation ($p_0=0.033$, $\rho=0.36$ and $p_0=0.05$, $\rho=0.33$ respectively). This shows that the correlation is stronger when we look only at the images with higher resolution and then suggests that the absence of correlation for the low spatial resolution is due to some bias. 
The same effect is measured with the magnification and the signal to noise ratio which are both strongly correlated with the image-plane resolution. Finally the different models used to reproduce the LLAMAS galaxies, of which 7 out of 9 are completely independent of UV spatial properties could probably play a role in measuring a lower correlation between Lyman-$\alpha$ and UV spatial extent.\\

\begin{figure*}
    \begin{minipage}{18cm}
    \includegraphics[width=18cm]{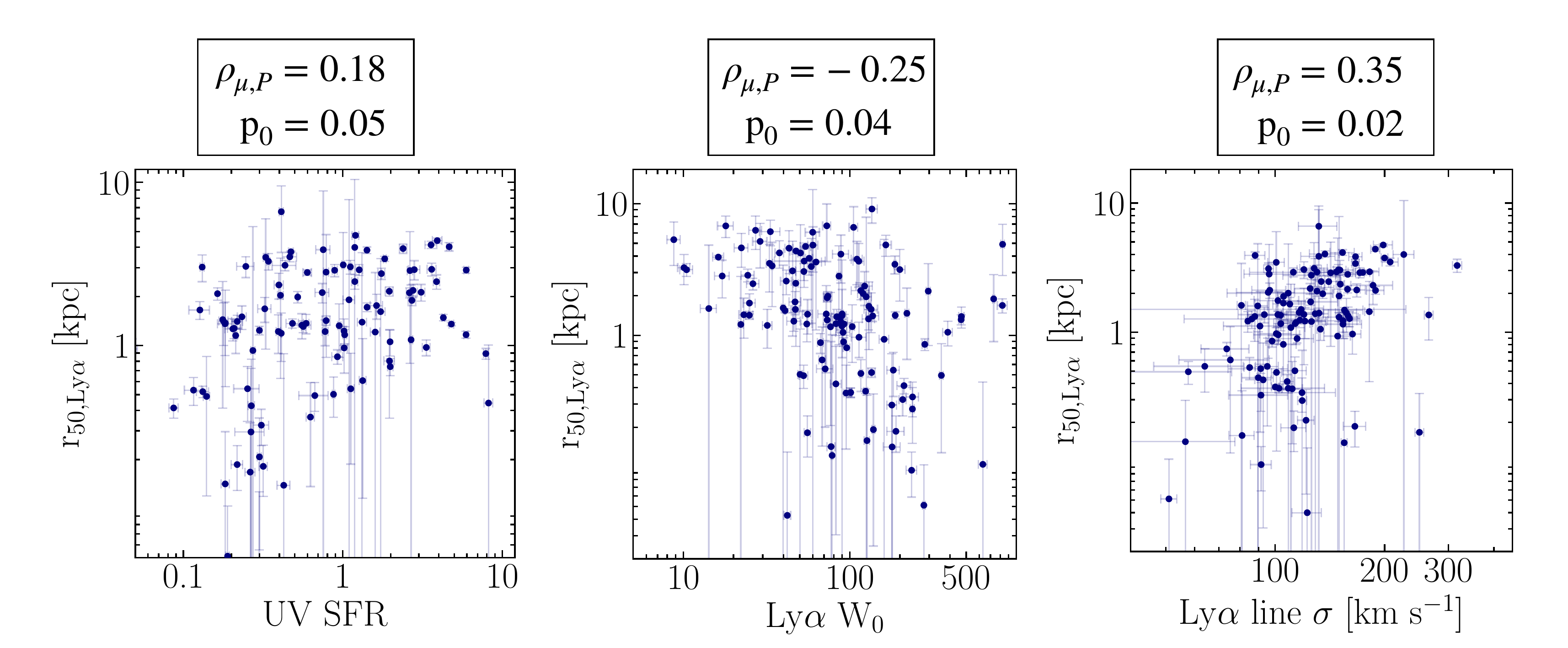}
    \end{minipage}
    \caption{$\rm r_{50,Ly\alpha}$ with respect to, from left to right, the UV SFR, Lyman-$\alpha$ rest-frame equivalent width ($W_0$) and the velocity dispersion of the global Lyman-$\alpha$ line ($\sigma$) on the x-axis. We indicate weighted Pearson coefficient $\rho_{\mu}$ and the associated p-value $p_0$ on top of each panel. }
    \label{fig:extended_corr}
\end{figure*}

We searched for correlations between the Lyman-$\alpha$ emission size (using the $r_{50, Ly\alpha}$ parameter) and the physical parameters governing the host galaxies (star formation rate, Lyman-$\alpha$ equivalent width, flux and luminosity, and properties of the Lyman-$\alpha$ line profile), and find three more or less significant correlations presented in Figure~\ref{fig:extended_corr}. First we find a very significant positive correlation between $r_{50,Ly\alpha}$ and the FWHM of the Lyman-$\alpha$ line (with weighted Pearson coefficient of $\rho=0.35$ and p-value of $p_0=0.0002$). Secondly, we observe a negative trend between $r_{50,Ly\alpha}$  and the Lyman-$\alpha$ rest-frame equivalent width $W_0$ (with weighted Pearson coefficient of $\rho=-0.25$ and p-value of $p_0=0.07$; the more extended the Lyman-$\alpha$ halo, the smaller is $W_0$. Finally we observe a weaker positive correlation between $r_{50,Ly\alpha}$ and the UV SFR ($\rho=0.18$ and $p_0=0.0002$). We observe similar correlations between theses three parameters and $r_{90,Ly\alpha}$.

\subsection{Spatial offsets between UV and Lyman-$\alpha$ emissions}
\label{sec:offset}
\begin{figure*}
    \includegraphics[width=18cm]{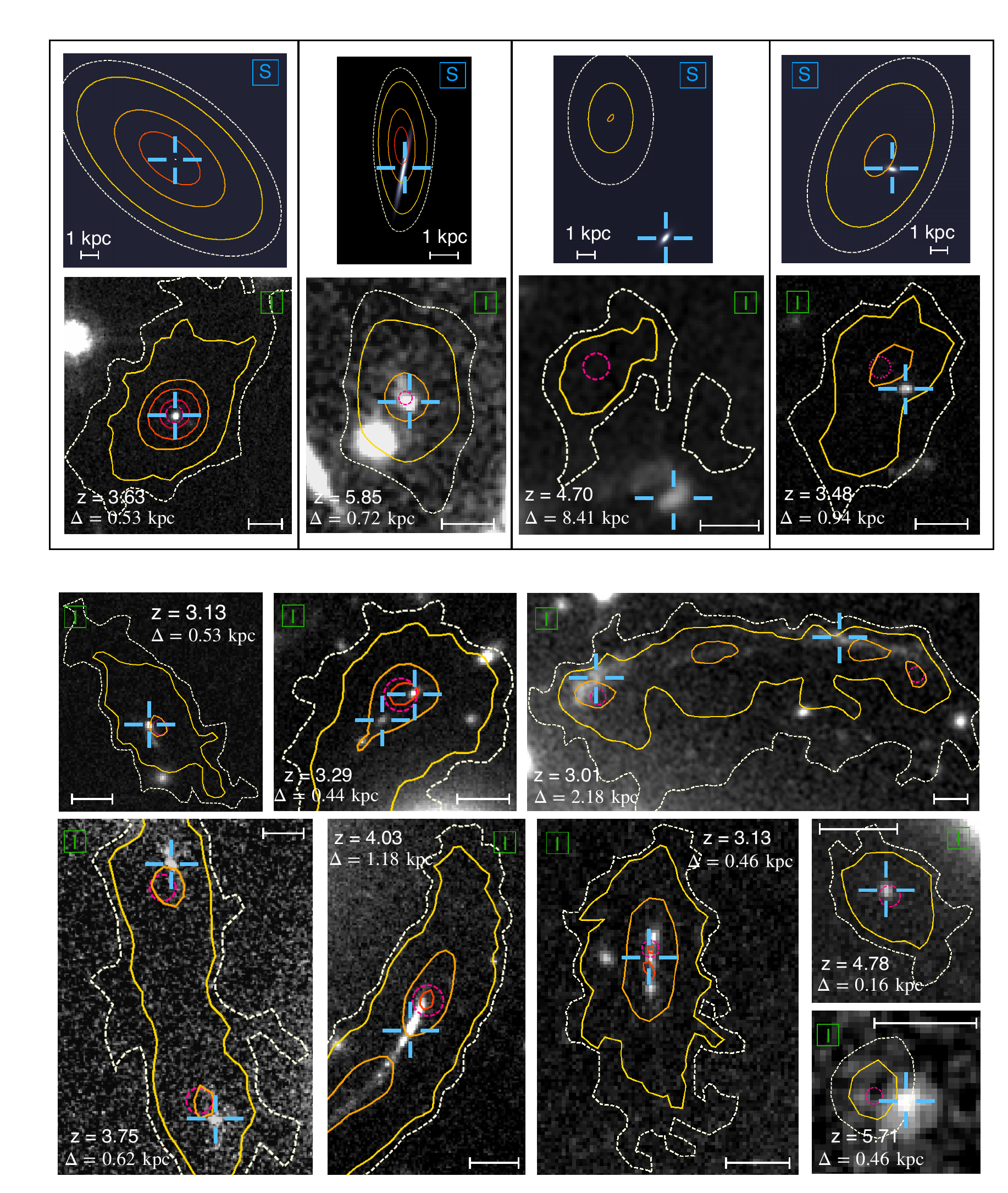}
    \caption{Presentation of 12 objects from the LLAMA sample presenting different values of spatial offsets. For the 4 galaxies on the top row, we show also the source-plane reconstruction. The observed images are labelled with a green 'I' and the source-plane modelisations with a blue 'S'. 
    In each panel, we show HST images (F814W or F606W depending of the redshift) or source-plane reconstruction. The MUSE Lyman-$\alpha$ emission is represented by the red, orange, yellow and white contours which correspond at a smooth surface brightness level of $250$, $100$, $50$ and $12.5 \ \rm 10^{-19} \ \rm erg.s^{-1}.cm^{-2}.arcsec^{-2}$, respectively. The ID, redshift and magnification values are given for each object. The $\Delta$ value is the value of the spatial offset measured  in the source plane between the UV and Lyman-$\alpha$ best model centroids. The blue cross indicates the UV position in the image plane and the pink circle the Lyman-$\alpha$ one. In all panels we add a scale (white line) of 1" in the image-plane and of 1 kpc in the 4 source-planes images .}
    \label{fig:offsets_im}
\end{figure*}

\begin{figure}
    \begin{minipage}{9cm}
    \includegraphics[width=9cm]{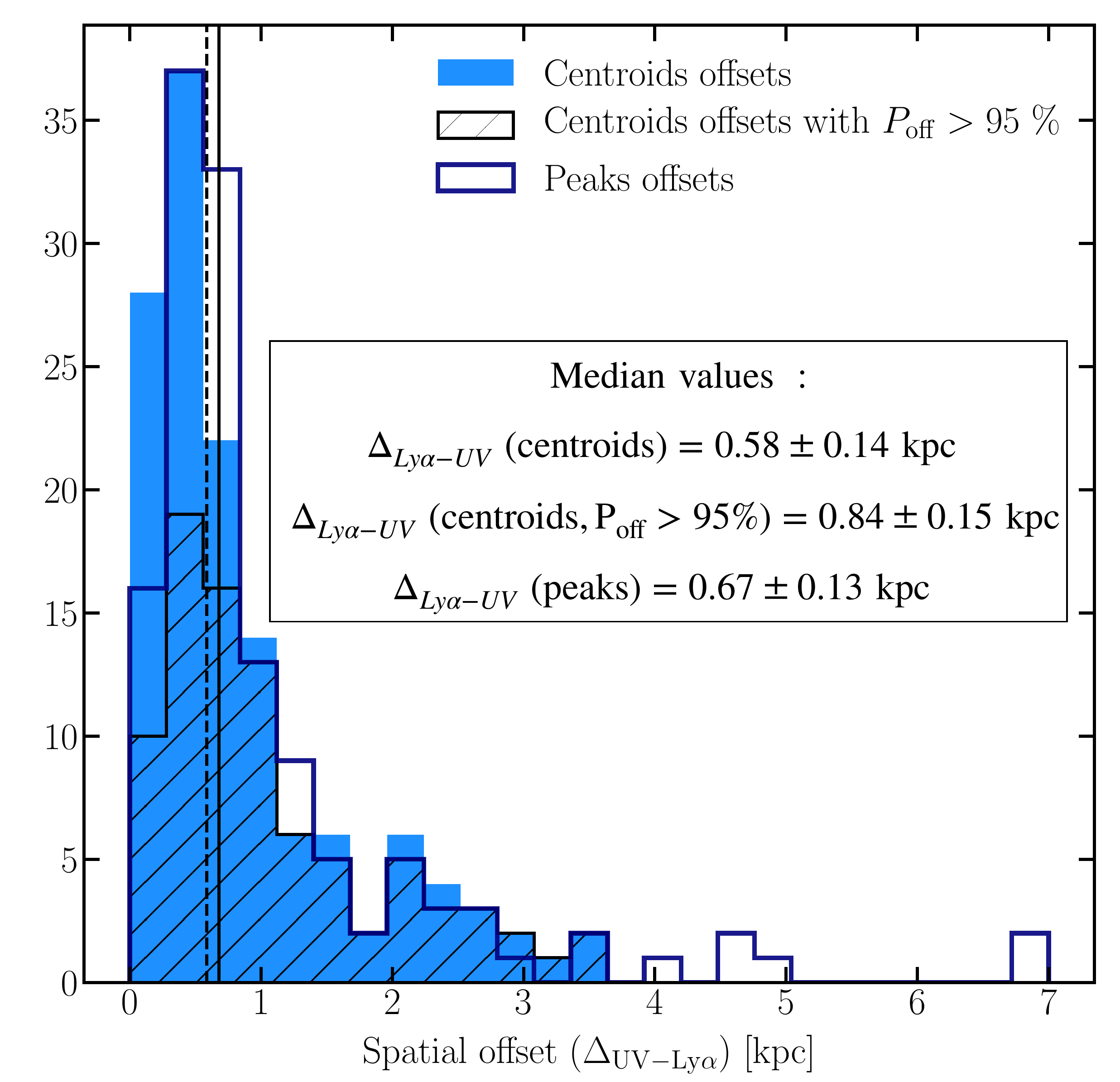}
    \end{minipage}
    \caption{Distributions of measured offsets in kpc in the source plane. The light blue distribution represents spatial offsets measured between UV and Lyman-$\alpha$ emission centroids measured in the source plane, and the dark blue one the distribution of spatial offsets measured between UV and Lyman-$\alpha$ emission peaks projected in the source plane. The hatched distribution represents the galaxies with a spatial offset probability higher than 95\%. The dashed and solid lines show the median values of centroids and peaks offsets distributions respectively.  }
    \label{fig:offsets}
\end{figure}

Looking at the best model of each UV and Lyman-$\alpha$ source, we notice a significant spatial offset between the two peaks locations; some examples are shown in Fig~\ref{fig:offsets_im}.
We measure this offset in each galaxy with a UV component observed in HST and selected for spatial fitting (see Section 3.4). We performed two different measurements on each galaxy. First we measured the two locations of peak emission (UV and Lyman-$\alpha$) in the image plane. We measure the peak location after a spatial Gaussian filtering of FWHM=0.2 or 0.1" for Lyman-$\alpha$ NB and HST images, respectively. We ray-trace these values into the source plane to measure the intrinsic (physical) spatial offset between the two emission peaks, in kpc. The distribution of this "peak to peak" offsets measurements is presented in dark blue in Figure~\ref{fig:offsets}. For the total sample, we measure a median value of $\rm \Delta (peaks) = 0.67 \pm 0.13$ kpc. When we keep only spatial offsets larger than 2 MUSE pixels  (i.e. 0.4") in the image plane, as they are considered more significant, the median value became $\rm \Delta (peaks) = 1.16$ kpc.
Second, we measure the spatial offset between UV and Lyman-$\alpha$ best model centroids (measured on the best-model source-plane image). For models with 2 Lyman-$\alpha$ emission peaks (M7 and M8 in Table~2), we measure the offset between the UV component centroid and the 2 Lyman-$\alpha$ peaks and kept only the smallest. 
The distribution of these offset measurements is shown in light blue in Figure~\ref{fig:offsets}, with the median value of the total sample is $\rm \Delta ({ centroids}) = 0.58 \pm 0.14$ kpc.  
Among the total sample, we identified 3 galaxies (two of them presented in  Fig~\ref{fig:offsets_im}) which present a strong Lyman-$\alpha$ absorption feature (LBG) in their spectrum and a huge spatial offset between UV and Lyman-$\alpha$ component. For these three sources we can attribute the large offset value to the absorption of Lyman-$\alpha$ at the location of the UV component. 
We compare our results with the three recent studies on spatial offsets performed by \citet{Hoag2019, Lemaux2020} and \citet{Ribeiro2020} on three samples with similar ranges of Lyman-$\alpha$ luminosities and UV magnitudes, and found a very good agreement. \citet{Hoag2019} measured the spatial offsets in 300 galaxies at $3<z<5.5$ observed in slit data, and find a median value of $0.61 \pm 0.05$ kpc. \citet{Ribeiro2020} measured a similar value in a sample of $\sim900$ galaxies at $2<z<6$ of $0.60\pm 0.05$ kpc. When they selected 11\% of galaxies with  secure offsets (after a visual inspection), the median value become $1.9 \pm 0.2$ kpc. Finally \citet{Lemaux2020} measured for 64 objects (mix of lensed and non-lensed galaxies) with $5<z<7$ a median offset of  $0.61 \pm 0.05$ kpc. In LLAMAS galaxies, we do not measure any significant variation of the offset values with redshift.
\\
Finally, we measure the impact of the lensing model to the offset measurements (see Figure~\ref{fig:displensing}). We found that 75\% of the multiple systems have a dispersion smaller than the mean value.

\subsection{Significance of the offsets measurements}
\label{sec:4.4}
To measure the robustness of these measurements we estimate the probability of measuring such offsets if Lyman-$\alpha$ and UV peaks are supperposed. Using the MCMC optimisation result of the best model from {\sc lenstool}, which provides a list of values of $x$ and $y$ positions in the source plane tested during the optimisation, for both UV and Lyman-$\alpha$ models. We apply the measured offset  to the Lyman-$\alpha$ position sample to center the cloud of Lyman-$\alpha$ positions on the UV ones. We randomly draw 10.000 pairs of UV and Lyman-$\alpha$ positions and measured for each couple the offset value $\delta_{95}$, measured in this way, corresponding to a cumulative probability to randomly measure an offset smaller than the real one $\Delta$ of 95\%. In other words it corresponds to the offset value from which there is less than 5\% of chance to randomly measure a similar or larger offset from the best models. We consider that an offset measurement is significant when $\delta_{\rm 95}<\Delta$. We find that 63\% of galaxies have a significant offset. Among the remaining 37\%, we measure with the same method, if the offsets corresponding to the cumulative probability of 68\% ($\delta_{\rm 68}$ at one sigma) which are smaller than $\Delta$ (corresponding to offset significant at only one sigma). We consider that all the galaxies with $\Delta<\delta_{\rm 68}$ are compatible with the non-offset scenario. For these galaxies, we use the $\delta_{\rm 68}$ as an upper limit to the offset measurement.
We measure this fraction of spatial offsets in 3 bins with $0<\Delta<1$ kpc, $1<\Delta<2$ kpc and $\Delta>2$ kpc. We find respectively 48\%, 92\% and 92\% of significant offsets. The fraction of measurements compatible with the non-offset hypothesis is of 37\%, 5\% and 2\% in the 3 same bins, respectively. Note that non significant offsets could simply be due to small intrinsic values or indeed coincident UV and Lyman-$\alpha$ emission. \\
\\
The spatial offsets measured in kpc in the source plane should be correlated to the UV size of the galaxy to be physically interpreted. With the aim to propose a more consistent measurement of spatial offset with respect to the UV continuum emission size, we measure the elliptical distance, $\Delta_{\rm ell}$ (normalised by the UV emission 90\% isocontour elliptical parameters) between the Lyman-$\alpha$ emission centroids and the UV emission centroid:
\begin{equation}
\label{eq:delta_ell}
    \Delta_{\rm ell}=\Big(\frac{X_{Ly\alpha}}{R_{\rm x, UV}}\Big)^{2}+\Big(\frac{Y_{Ly\alpha}}{R_{\rm y,UV}}\Big)^2
\end{equation}
with ($X_{Ly\alpha},Y_{Ly\alpha}$) the position of the Lyman-$\alpha$ emission centroid in the referential formed by the axis of the UV emission elliptical distribution and ($R_{\rm x,UV}, R_{\rm y,UV}$) respectively the semi-major and semi-minor axis of the UV elliptical emission distribution.

\begin{figure}
    \begin{minipage}{9cm}
    \includegraphics[width=9cm]{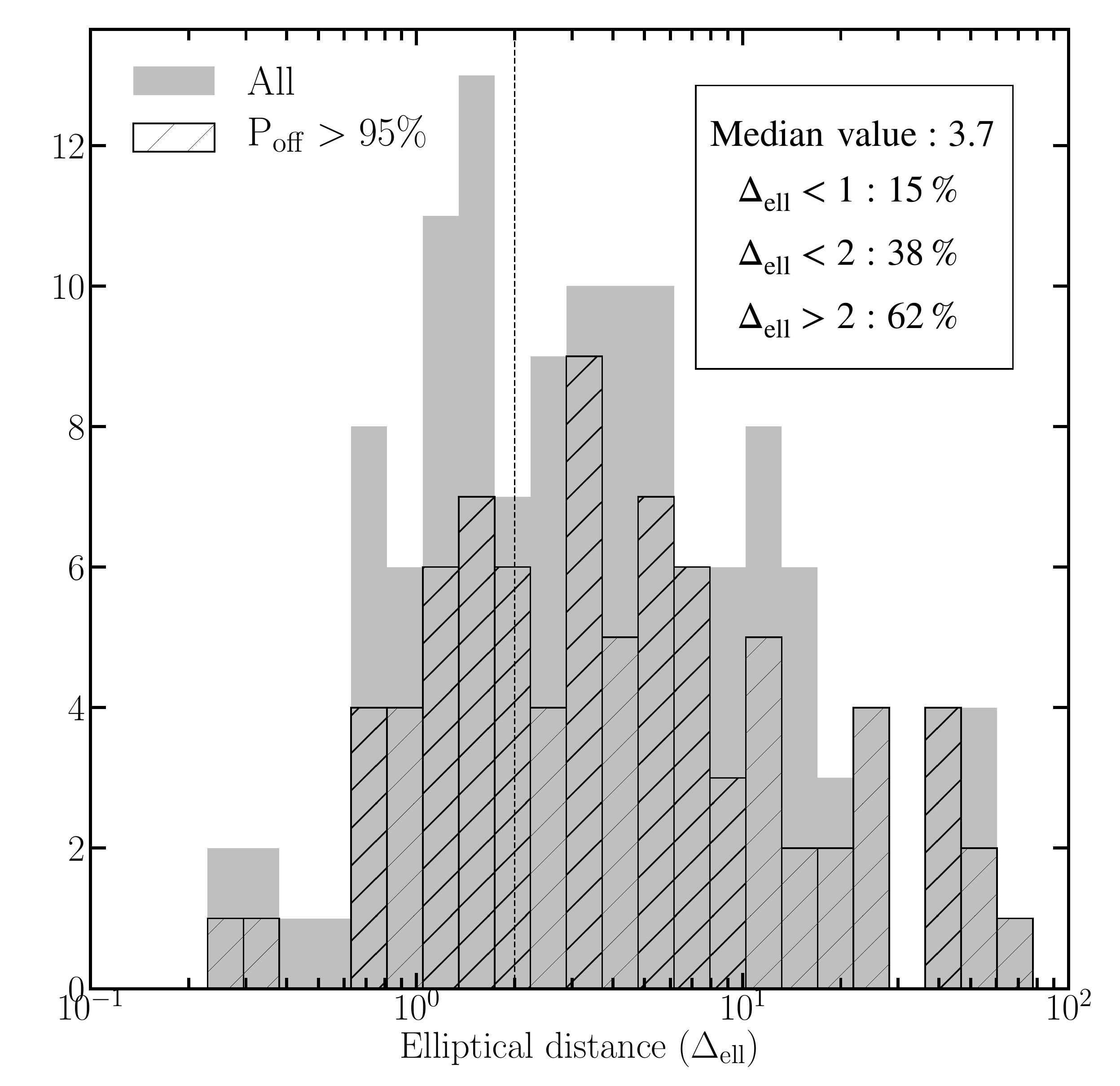}
    \end{minipage}
    \caption{Distribution of elliptical distance (cf Section~\ref{sec:4.4}) measured in the source plane between Lyman-$\alpha$ emission centroids and the ellipse formed by the UV emission distribution (using $\rm r_{90,UV}$ as radius). The grey distribution represents all the  LLAMAS galaxies and the black hatched only the galaxies with a spatial offset probability higher than 95\%. The vertical dotted line represents the separation between internal and external spatial offsets ($\Delta_{\rm ell}=2$). }
    \label{fig:r_ell}
\end{figure}

A value of $\Delta_{\rm ell}<1$ means that the Lyman-$\alpha$ emission center is located in the UV continuum component and probably produced by substructures within the star-forming region. When $\Delta_{\rm ell}>1$, the Lyman-$\alpha$ emission peak is produced outside of the stellar component, probably by satellite galaxies, or large-scale effects in the CGM. 
We show the distribution of $\Delta_{\rm ell}$ values in grey in Fig~\ref{fig:r_ell}. We measure that 40\% of the galaxies present an external spatial offset as high as  $\Delta_{\rm ell}$>2. \\

We measure no correlation (p-values > 0.05) between the values of elliptical distance and the physical parameters of the host galaxies (Ly$\alpha$ $W_0$, flux and luminosity, $\Delta$ in kpc). After dividing their sample of galaxies in 2 bins, based of $W_0$, \citet{Hoag2019} found a significant trend: galaxies with lower $W_0$ values  show higher offset values (mean of $1.92 \pm 0.13$ kpc) than galaxies with higher $W_0$ (mean of $1.51\pm 0.11$  kpc). We do not confirm this trend, neither with offset value in kpc nor with the elliptical distance. \citet{Lemaux2020} observed that the spatial offset increases with the UV brightness of the galaxies. We find the opposite correlation when we compare UV SFR values and elliptical distances for all galaxies with secure spatial offset measurements (90 sources). Galaxies with an elliptical distance $\Delta_{ell}<3.9$ (45 sources) show on average a higher UV SFR value ($\langle SFR \rangle=1.74 \pm 0.23 \ \rm M_\odot\ yr^{-1}$) than galaxies with $\Delta_{ell}>3.9$ (45 sources, $\langle SFR \rangle =1.26 \pm 0.28 \ \rm M_\odot\ yr^{-1}$). We notice also that the galaxies with a higher elliptical distance present a smaller UV size ($\langle r_{\rm 50, UV} \rangle =0.08 \pm 0.01$ kpc) and larger Lyman-$\alpha$ versus UV emission extent ($\langle\frac{ r_{\rm 50, Ly\alpha}}{ r_{\rm 50, UV}}\rangle=25.5 \pm 4.3$), than galaxies with smaller elliptical distances ($\langle r_{\rm 50, UV}\rangle =0.31 \pm 0.4$ kpc and $\langle\frac{ r_{\rm 50, Ly\alpha}}{ r_{\rm 50, UV}}\rangle =7.0 \pm 0.8$). Thus, more than half of the Lyman-$\alpha$ peaks are located outside of the stellar body of the source which could be due to the presence of an extra Lyman-$\alpha$ emission source such as a satellite or merging galaxy.

Besides, it should be easier to distinguish a spatial offset in the Lyman-$\alpha$ emission produced by a satellite galaxy for smaller (low $r_{\rm UV}$) galaxies, because these sources should emit less Lyman-$\alpha$ photons and thus the contribution of a satellite galaxy  would be easily detectable in the global profile. This could explain why the galaxies for which we measured higher elliptical distances present on average a smaller UV component.


%--------------------------------------------------------------------

\section{Discussion}
\label{sec:discussion}
\begin{figure*}
    \includegraphics[width=18cm]{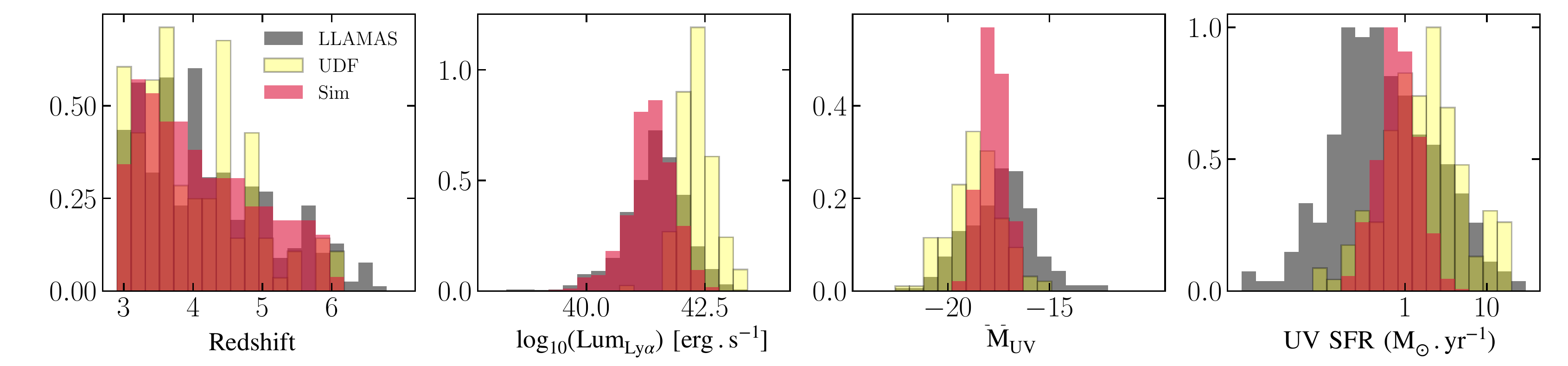}
    \caption{Global physical properties of the simulated, UDF and LLAMAS LAEs. From left to right: redshift, Lyman-$\alpha$ luminositie, UV magnitude and SFR distributions. The Lyman-$\alpha$ luminosity and UV magnitude of the LLAMAS galaxies are intrinsic values measured (i.e. unlensed values). LLAMAS galaxies represented here are only the objects selected for spatial fitting (see Sect.~\ref{sec:LAE_selection}). All values are measured following the same procedure on UV and NB Lyman-$\alpha$ images.}
    \label{fig:hist_simus}
\end{figure*}
%\subsection{Predictions from hydrodynamical simulations}

\subsection{Production of mock observations}
In order to physically interpret these results we compare the measured LAE properties with a  cosmological radiation hydrodynamical (RHD) simulation of a high-redshift galaxy evolving from $z=6$ to $z=3$, described in \citet{Valentin}. This zoom-in simulation was produced using the {\sc ramses-rt} code (\citealt{Rosdahl2013,Rosdahl_Teyssier}). The simulation includes all the expected Lyman-$\alpha$ production mechanisms (photo-ionization and photo-heating of hydrogen by local sources, collisional excitation of hydrogen as well as contribution from the UV background) and so represents a powerful tool to study Lyman-$\alpha$ photons escape in both ISM and CGM. This simulated galaxy has been deliberately chosen to be representative of the faint-UV LAEs detected in the recent MUSE studies of LAEs at high redshift (\citealt{Leclercq2017, Lutz2016}) in terms of halo mass ($M_{h}=6 \times 10^{10} \rm M_{\odot}$ at $z=3$). We study mock observations of the Lyman-$\alpha$ line cubes and UV continuum emission at 1500 \AA \ rest-frame. Each mock Lyman-$\alpha$ dataset consists in a 10"$\times$10"$\times$10.9 \AA \ datacube centered on the Lyman-$\alpha$ line with 0.067"$\times$0.067"$\times$0.0625 \AA\ pixels, which is three times better than the MUSE sampling both in spatial and spectral directions. The UV continuum 1500 \AA \ rest-frame images are produced with 0.01"$\times$0.01" pixels which are 5 times smaller than HST/ACS pixels. 
To represent the diversity of observed Lyman-$\alpha$ profiles, 12 mocks datacubes were produced for each of the 129 simulation timesteps, by projecting along 12 different line of sights defined by healpix Nside=1 (identical at all redshifts). Thus 12$\times$129 mocks were produced at the 129 different redshifts ranging from $z=3.000$ to $z=5.989$ with a regular lookback time interval of 10 Myrs. Although this simulation focuses only on one galaxy, the fact that this galaxy is studied at 129 different redshifts in 12 different directions adds some diversity due to variations in SFR with time, galaxy growth, effects of radiative transfer into the CGM and line of sight projections. The global properties of the simulated galaxy sample and UDF and LLAMAS sample are presented in Figure~\ref{fig:hist_simus}. Altogether, this sample of mock observations has physical properties close to the LLAMAS properties in terms of redshift, Lyman-$\alpha$ luminosity and SFR.
The median redshift value is $z=3.92$ for the simulated data versus $z=4$ for the LLAMAS sample. Note that the distribution of UV magnitudes in the simulation is narrower than that of LLAMAS galaxies, but it roughly covers a similar range of M$_{\rm UV}$ with a median value slightly brighter than the median value of the LLAMAS sample ($-17.8$ vs. $-17.0$ respectively).

The SFR of the simulated galaxy varies with redshift, on average the SFR increases during the formation and evolution of the galaxy from $0.5 \ M_{\odot}.yr^{-1}$ to $1.23 \ \rm  M_{\odot}.yr^{-1}$ between $z=6$ and $z=3$ with few SFR peak episodes (up to $3.0 \ \rm  M_{\odot}.yr^{-1}$) during its history. These SFR variations are strongly correlated with the variation of the total Lyman-$\alpha$ luminosity. The median value of the SFR on the simulated sample ($\rm SFR_{\rm Sim}= 0.80 \  \rm M_\odot .yr^{-1}$) is larger than the LLAMAS median value ($\rm SFR_{\rm LLAMAS}= 0.48 \ \rm M_\odot.yr^{-1}$), although lower and higher SFRs can also be found in the LLAMA sample.
We notice a very good match in terms of Lyman-$\alpha$ luminosities between LLAMAS and the simulation, with Lyman-$\alpha$ luminosities ranging from $\rm log(Ly\alpha / \rm erg.s^{-1}) \sim 40$ to $42.5$. with a median value around $41.5$.

We note that the UV size of the simulated galaxy increases from $0.15$ to $0.45$ kpc between $z=6$ and $z=3$ as a result of the continuous mass growth due to gas accretion and mergers over this period of $\sim$ 2 Gyr.

To compare with lensed and non-lensed MUSE observations of high redshift LAEs, we produced mock "UDF-like" and "LLAMAS-like" Lyman-$\alpha$ NB and HST images from the simulated galaxies. To produce Lyman-$\alpha$ NB images we collapse a cube containing all the Lyman-$\alpha$ emission without continuum.
To reproduce non-lensed MUSE Lyman-$\alpha$ NB images of LAEs, we convolve the initial raw Lyman-$\alpha$ NB images by a typical MUSE UDF PSF, depending on the redshift of the galaxy, as described in Section ~\ref{sec:PSF}. Finally we re-grid the images at the MUSE sampling of 0.2"$\times$0.2" pixels and then add a random Gaussian noise based on the typical level observed in UDF Lyman-$\alpha$ NB images.
We follow the same method for UV images; we convolve the raw UV images with UDF HST PSF, constructed following the method explained in Section~\ref{sec:PSF}, using HST F606W or F814W images (depending on the redshift).  We finally re-grid the images at the HST sampling of 0.05"$\times$0.05". We add a random Gaussian noise based on the typical level observed in the HST images of the UDF.
To reproduce "LLAMAS-like" observations, we first chose randomly a cluster model and a specific source location from the R21 data release. We lens both UV and Lyman-$\alpha$ NB images by the lens model. We then follow the same procedure for PSF convolution, pixels sampling and addition of Gaussian noise as for UDF-like observations. We use typical MUSE PSF parameters and HST PSF measured in the cluster used as a lens.
For both kinds of observations (UDF-like and LLAMAS-like) we use the python package {\sc photutils} to detect the different images, following the same criteria used for the LLAMAS sample (see section ~\ref{sec:LAE_selection}). When the lensing produces multiple images, we keep only one of them for comparison with observations. We apply a S/N threshold of S/N=6 for Lyman-$\alpha$ NB "UDF-like" images as applied in \citet{Leclercq2017}. For LLAMAS-like images, we apply the same selection for spatial fits as for real observations, highlighted by the blue contour in Figure~\ref{fig:select_llamas_flux}. 
We finally obtain a sample of 1164 raw images (both UV and Lyman-$\alpha$ NB), 271 "UDF-like" images and 254 "LLAMAS-like" (the other produced UDF-like and LLAMAS-like images had too faint UV or Lyman-$\alpha$ images to be detected or spatially characterised).
    
\subsection{Extended Lyman-$\alpha$ haloes}

\subsubsection{Variations in Lyman-$\alpha$ extent}

\begin{figure}
    \begin{minipage}{9cm}
    \includegraphics[width=9cm]{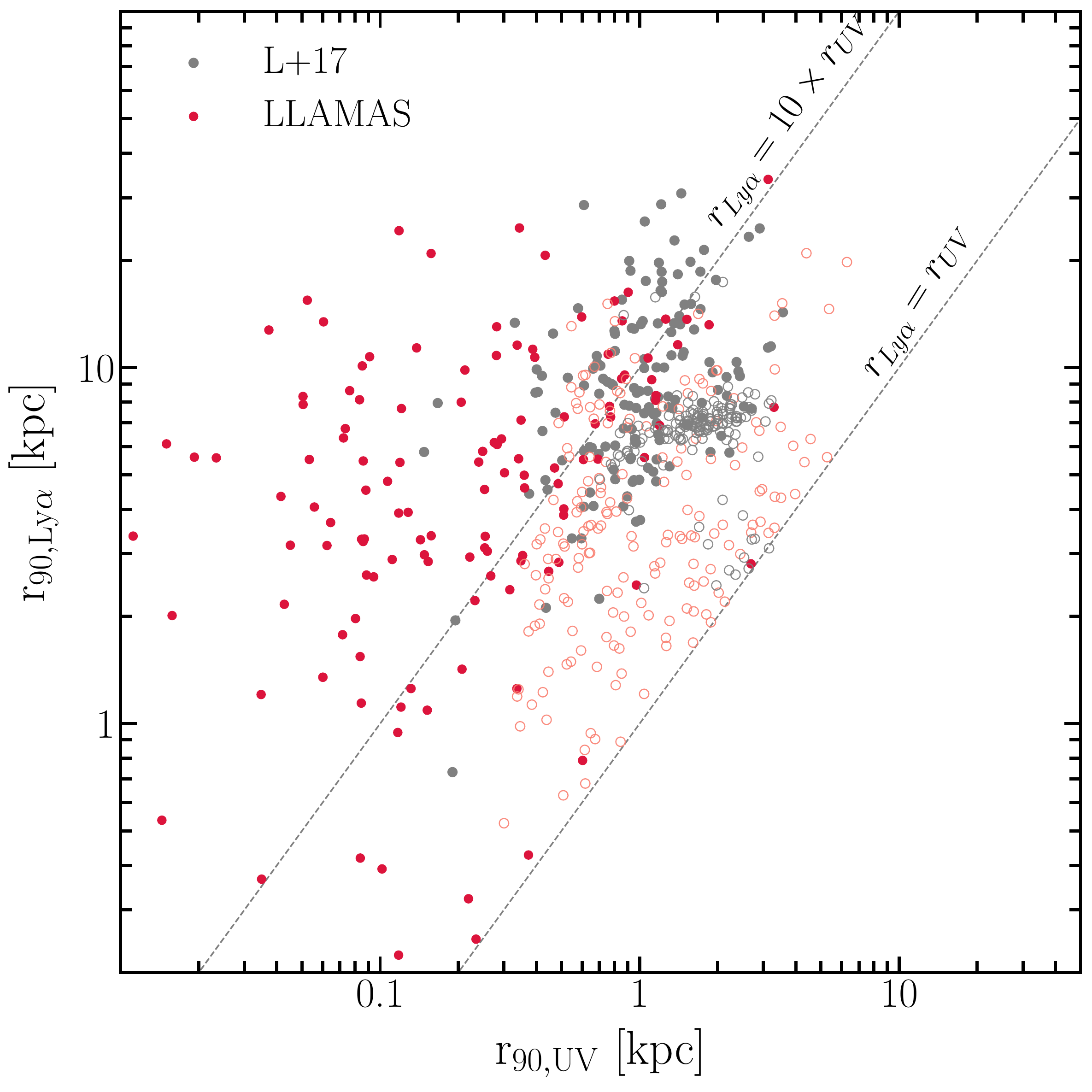}
    \end{minipage}
    \caption{Lyman-$\alpha$ emission 90-light radius $r_{90,Ly\alpha}$ as a function of the UV emission 90-light radius $r_{90,UV}$ for UDF galaxies (in grey), LLAMAS galaxies (in red), UDF simulated galaxies (empty grey circles) and LLAMAS simulated galaxies (empty red circles).}
    \label{fig:plot_jeremy}
\end{figure}

We apply the same spatial fit as presented in Section~\ref{sec:source_plane_fit} to the three types of mock observations ('raw' simulation, UDF-like and LLAMAS-like observations). For UV measurements, we apply a single elliptical exponential component fit on the three different mock observations. For the raw data and UDF-like observations we apply on Lyman-$\alpha$ NB images a single elliptical exponential two components fit (M4 in Table~\ref{tab:fits_description}). For LLAMAS-like mock observations, in order to fairly compare with the LLAMAS sample results, we applied 3 different fits on each selected image and compared BIC criteria to choose the best model of each galaxy. We choose to apply the models labelled M4, M6 and M8 in Table~\ref{tab:fits_description}. We measured half-light and 90\%-light radius on best-fit images for each source. The morphology of simulated LLAMAS haloes gives a similar range of concentrations ($c=1.16\textrm{--}32.5$) as the real observations.
Figure~\ref{fig:plot_jeremy} shows the distribution of Lyman-$\alpha$ $\rm r_{90}$ for UDF, LLAMAS, UDF-like and LLAMAS-like galaxies. We observe that the simulated UDF galaxies (empty grey points) are confined to a small region of the cloud of points from \citet{Leclercq2017}. The simulated LLAMAS-like galaxies are located in a different region, presenting both smaller and higher values of Lyman-$\alpha$ extent, closer to the real LLAMAS galaxy Lyman-$\alpha$ spatial extent values. In all LLAMAS-like and UDF-like galaxies, the Lyman-$\alpha$ emission is measured to be more extended than the UV central component, as for real LLAMAS galaxies. However the mean $x=r_{90,Ly\alpha}/r_{90,UV}$ ratio is smaller for the simulated galaxies whatever the simulated observation method ($\langle x_{\rm 90,UDF-sim} \rangle =4.3 $ and $\langle x_{\rm 90,LLAMAS-Sim} \rangle = 5.0 $). On average the simulated LAEs present a Lyman-$\alpha$ emission which is less extended, with respect to the UV component, than the observed ones.
The fact that the same simulated galaxy leads to different physical parameters is indicative of an inconsistency between the two measurements which could be due as much to the measurement method as to the instrumental effects (PSF and noise).\\

\begin{figure}
    \includegraphics[width=9cm]{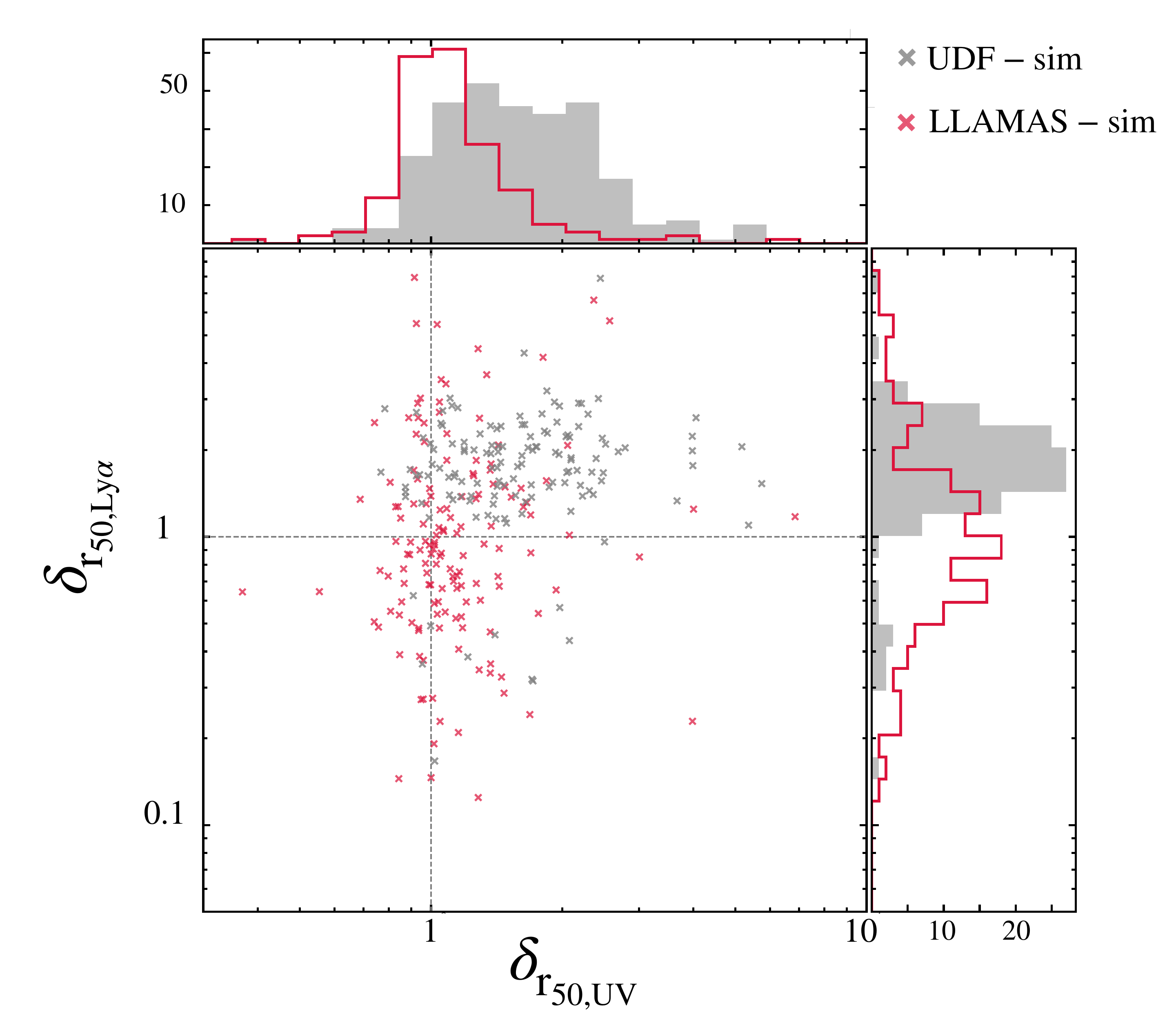}
    \caption{Distribution of the ratio of the measurements of $r_{50,Ly\alpha}$ and $r_{50,UV}$ for UDF simulated (in grey) and LLAMAS simulated (in red) galaxies to the same measurements done on raw simulated data. The middle panel shows the distribution of the points for the 2 samples and the top and right panel  the distribution of each parameter. The grey lines indicate $\delta_{50,Ly\alpha}=1, \ \delta_{50,UV}=1$.}
    \label{fig:delta_extent}
\end{figure}

Figure \ref{fig:delta_extent} represents the relative errors (i.e. the ratio of LLAMAS-like or UDF-like measurements and original mock measurement) of UV and Lyman-$\alpha$ size measurements both for LLAMAS-like (in blue) and UDF-like sources (in green). We notice that UDF measurement tend to overestimate both UV and Lyman-$\alpha$ emission sizes. This effect could be due to the PSF smoothing, dominant in the UDF simulated images. The UV measurements on the LLAMAS-like simulated galaxies are in good agreement with the values estimated on the original high resolution images of the simulation. The Lyman-$\alpha$ extent is less constrained, however no systematic bias is observed. These different results confirm the gain from the lensed samples in the study of extended Lyman-$\alpha$ emission.\\

\subsubsection{Origin of the extended Lyman-$\alpha$ emission}

We notice in the simulations a strong effect of the line of sight on the spatial extent measurements but we did not measure any significant correlation between the Lyman-$\alpha$ spatial extent and the physical parameters of the galaxies in the simulation both for the UDF-like and LLAMAS-like samples. 
Both in the UDF and LLAMA samples, the Lyman-$\alpha$ emission is almost always more extended than the UV component traced by the UV rest-frame emission thus dominated by the young stellar population emission. From the results of this study we propose two possible scenario to explain this result. First the Lyman-$\alpha$ haloes could be due only to the scattering of the Lyman-$\alpha$ photons from the source emission to the outskirts of the halo. This scenario is supported  by strong correlation measured between $\rm r_{50, Ly\alpha}$ and the width of the Lyman-$\alpha$ line and presented in the Figure~\ref{fig:extended_corr}. Indeed, assuming that more extended Lyman-$\alpha$ haloes trace optically thick media where the number of Lyman-$\alpha$ scatterings is increased, we expect from theoretical studies that these halos will exhibit broader line profiles as a result of resonant scattering (e.g. \citealt{Verhamme2018})

Secondly, the correlation measured between $\rm r_{50, Ly\alpha}$ and the UV SFR (Figure~\ref{fig:extended_corr}) indicate that the spatial extent of the CGM may also depend on the UV stellar activity. Anisotropic outflows, as observed in star-forming galaxies at $2<z<6$ (\citealt{Steidel2010, Pelliccia2020,Ginolfi2020}) can be produced by stellar feedbacks. These outflows could push the gas and thus let Lyman-$\alpha$ photons diffuse further away from the galaxy center by decreasing the covering fraction of the hydrogen gas \citep{Lemaux2020}; causing the halo expansion. This scenario could also explain the asymmetric and anisotropic Lyman-$\alpha$ emission distribution noticed in some LLAMAS galaxies. \citet{Armin2021} measured the same trend at low redshift (45 galaxies at $z<0.24$) where the Lyman-$\alpha$ emission extent is correlated with the stellar mass and the star formation regions sizes.


\begin{figure*}
    \begin{minipage}{18cm}
    \includegraphics[width=18cm]{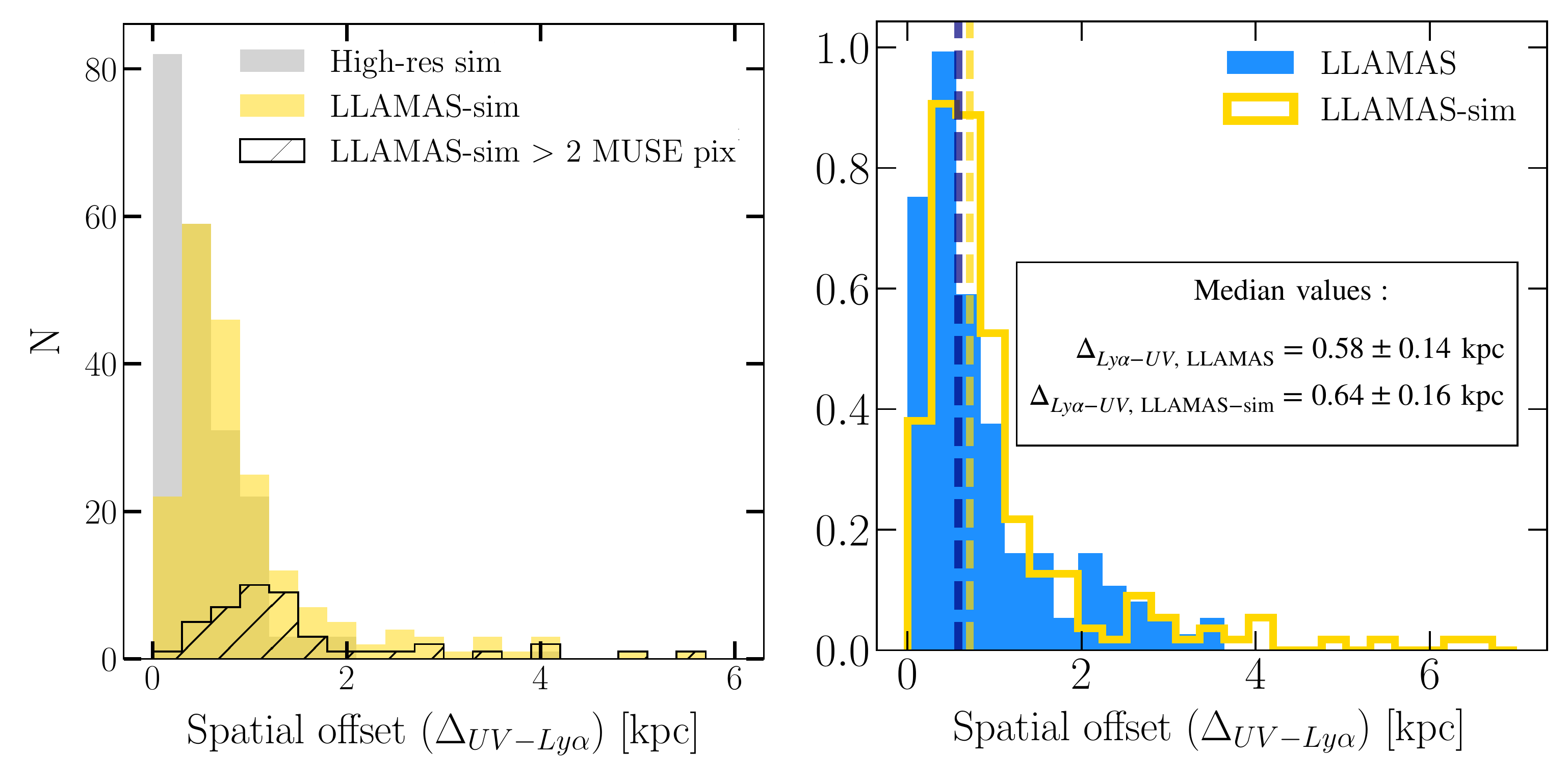}
    \end{minipage}
    \caption{Left: Distribution of the spatial offsets measured in the simulated raw data (grey), simulated lensed observations (yellow). The black distribution show the spatial offsets which are observed in the image plane higher than 2 MUSE pixels.  Right: Normalized spatial offset distribution for the LLAMAS galaxies (in blue) and the LLAMAS simulated galaxies (in yellow). The dashed lines show the mean value of the two samples. }
    \label{fig:off_simus}
\end{figure*}

%\subsection{Comparison of spatial offsets distributions}
\subsection{Spatial offsets between UV and Lyman-$\alpha$ emissions}

\subsubsection{Offsets in mocks vs. observations}
We measured a significant spatial offset between UV and Lyman-$\alpha$ emission in $60$\% of the LLAMAS galaxies, ranging from 0.1 to 7 kpc. Following the same procedure used in observational data (Sect. 3.2 ), we measured spatial offsets between UV and Lyman-$\alpha$ emission in the three samples of simulated sources. 
The left panel of Figure~\ref{fig:off_simus} shows the values of spatial offsets measured in raw simulations (grey) and LLAMAS-like sources (yellow). The values shown in orange are the sources with a spatial offset larger than 0.4" in the image plane (which is often the limit given in the literature to claim significant offsets in MUSE observations). We notice that a high number of LLAMAS-like sources (22\%) present an offset larger than 0.4" in the image plane, thanks to lensing magnification. The distribution of spatial offset measured on LLAMAS-like simulated galaxies is close to the raw simulations measurements. ($\Delta_{\rm  UV-Ly\alpha, simus}=0.40$ kpc and $\Delta_{\rm  UV-Ly\alpha, LLAMAS-like}=0.64$ kpc). In the UDF-like mocks, only 16 sources (6\%) show an offset larger than 0.4", which can explain why no spatial offsets are usually reported with MUSE in non-lensed galaxies due to resolution limits and emphasizes the gain provided by lensing surveys.
When we compare the distribution of spatial offsets measured in the LLAMAS sample with the distribution measured in the LLAMAS-like sample (right panel of Figure~\ref{fig:off_simus}), we find a very good match, highlighting that these simulated galaxies incorporate physical mechanisms capable of producing similar spatial offsets to those observed.
Figure~\ref{fig:r_ell} represents the values of $\Delta_{\rm ell}$ for the simulated LLAMAS-like galaxies in gold. We notice that the simulated galaxies have on average lower elliptical distances but span anyway a large range of $r_{\rm ell}$ values as measured in the LLAMAS galaxies. 72\% of the simulated LLAMAS-like galaxies present an internal spatial offset with $\Delta_{\rm ell}<2$. We do not measure very high values (>20) of $\Delta_{\rm ell}$ in the simulated galaxies. In the LLAMAS, we identify the galaxies with $\Delta_{\rm ell}>20$ as being LBGs with a strong absorption feature observed in their spectra (one example is presented in the Figure~\ref{fig:offsets} at $z=4.69$). 

\subsubsection{Origin of the spatial offsets in the simulation}

\begin{figure*}
    \begin{minipage}{18cm}
    \includegraphics[width=18cm]{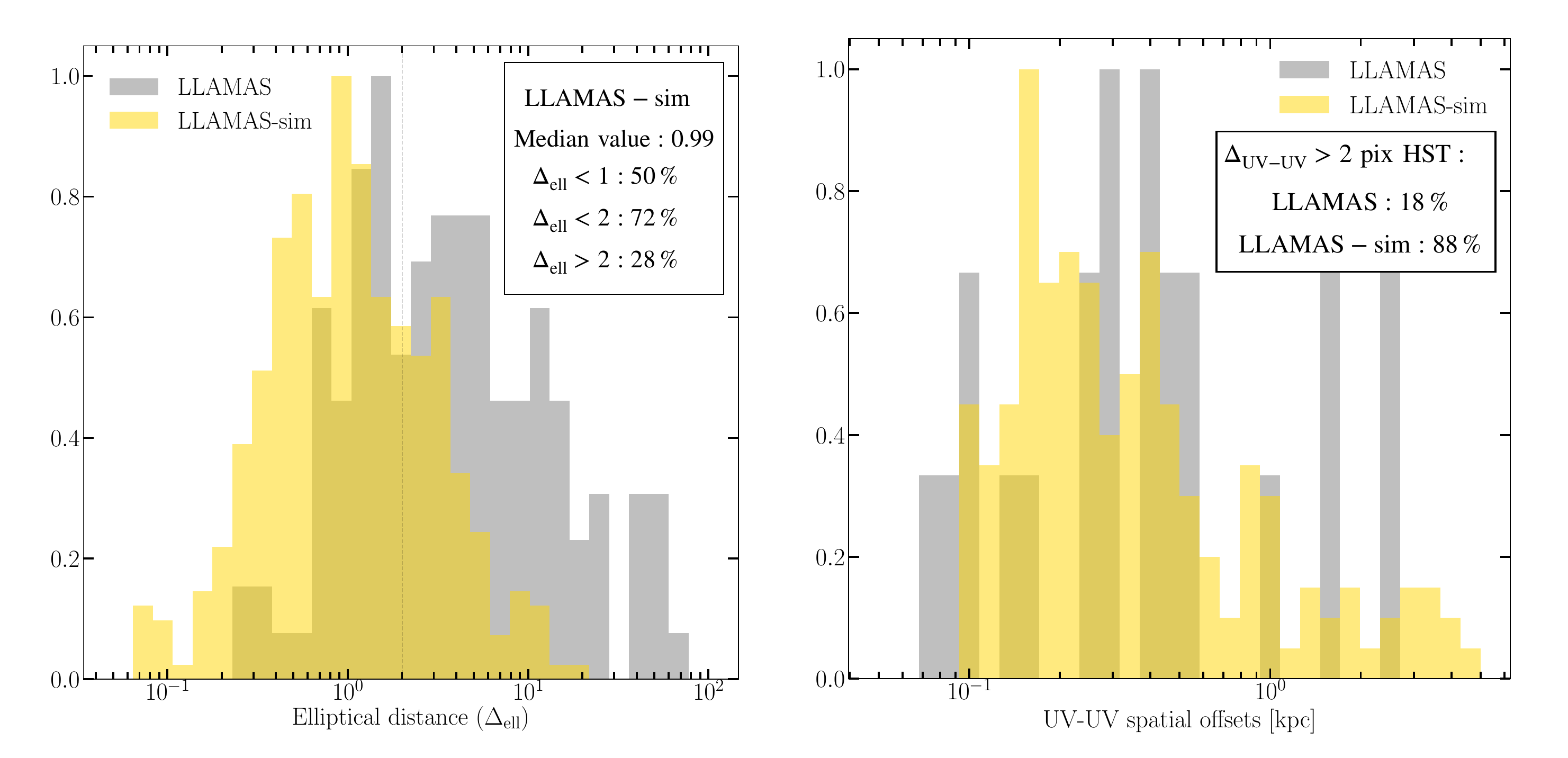}
    \end{minipage}
    \caption{Left: Distribution of elliptical distance measured in the source plane between the Lyman-$\alpha$ emission centroids and the ellipse formed by the UV emission distribution (using $\rm r_{90,\rm UV}$ as radius) for the LLAMAS galaxies (in grey) and the simulated LLAMAS galaxies (in yellow). Right: Distribution of the UV-UV spatial offsets measured between the UV emission peak and the UV centroid position in LLAMAS galaxies (in grey) and simulated LLAMAS galaxies in yellow. We show here only the galaxies for which this UV-UV offset is higher than 2 HST pixels in the image plane. }
    \label{fig:off2_simus}
\end{figure*}

\begin{figure*}
    \begin{minipage}{18cm}
    \includegraphics[width=18cm]{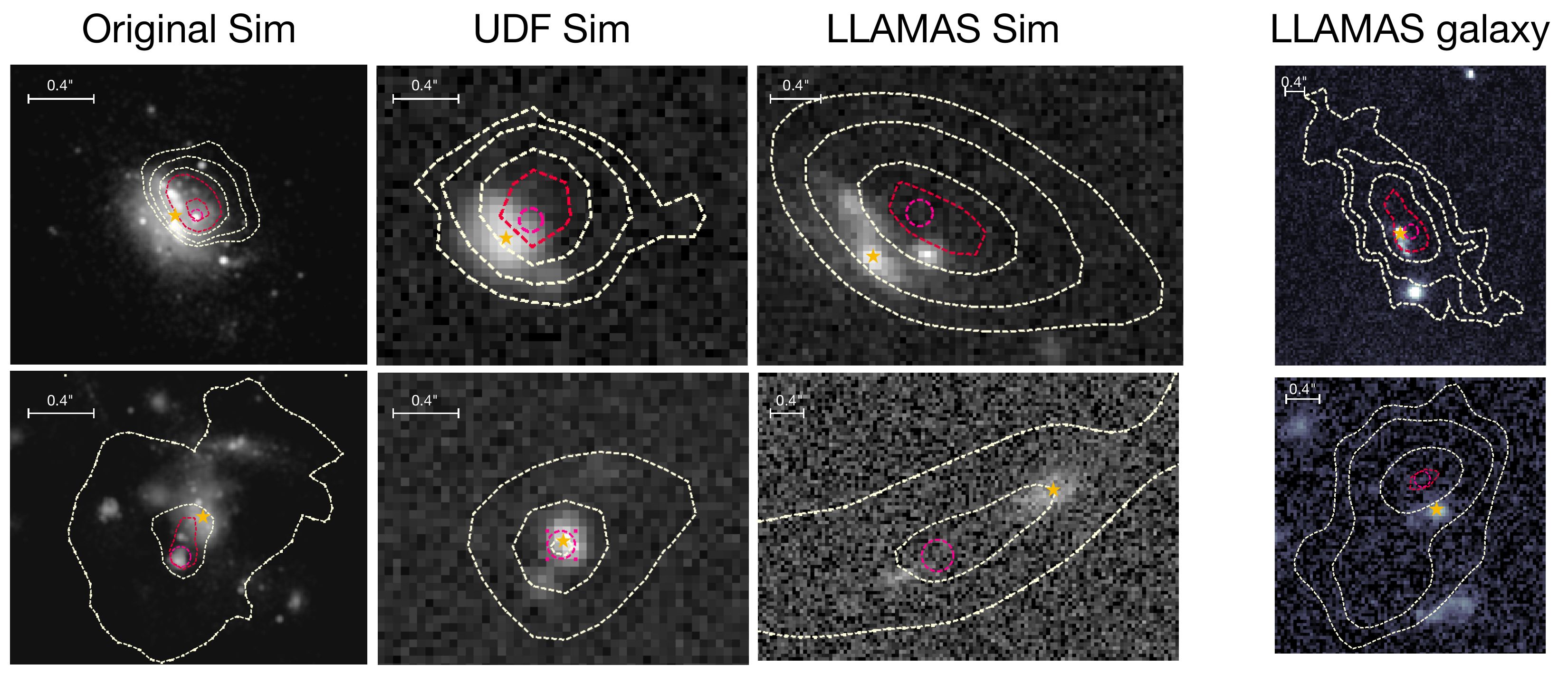}
    \end{minipage}
    \caption{Two examples of simulated galaxies. From left to right: high resolution image (UV emission), simulated HST image, simulated HST image of the same galaxy lensed by a LLAMAS cluster and HST image of a real LLAMAS galaxy. On each image, the contours represent the Lyman-$\alpha$ emission. The orange stars show the location of the centroid of the UV emission and the pink circle the location of the Lyman-$\alpha$ centroid. The first row presents an example of spatial offset produced by an offsetted UV bright clump. The second row presents an example of a spatial offset produced by a faint UV component spatially coincident with a strong Lyman-$\alpha$ emission peak.}
    \label{fig:ex_simus}
\end{figure*}

Thanks to the high spatial resolution of the simulation, we can investigate the origin of the spatial offsets found in the mocks.
Figure~\ref{fig:off2_simus} shows that there are two regimes in the elliptical distance distribution of LLAMAS galaxies: sources with $\rm \Delta_{ell} < 2$ and $\rm \Delta_{ell}>2$. 
Galaxies with  $\rm \Delta_{ell} <2$  represent 38\% \ of the LLAMAS sample and 72\% \ of the LLAMAS-like simulated galaxies. In this case, the spatial offset is likely due to an offsetted star formation clump emitting a high quantity of Lyman-$\alpha$ photons or due to the inhomogeneous neutral hydrogen distribution surrounding the galaxy. For low redshift galaxies, the LARS sample (\citealt{Hayes2013, Hayes2014, Ostlin2014}) observed a high distinct clumpiness of the ISM emission in both UV and H$\alpha$ emission as well as a complex structure of the Lyman-$\alpha$ emission.  As we know that such sub-structures are present in high redshift galaxies (\citealt{Elmegreen2013} and \citealt{Forster2018}), they could explain the formation of small offsets between Lyman-$\alpha$ and UV emission at the scale of the continuum component. In the high resolution images of the simulated galaxy, we can observe a very clumpy UV emission (see two examples in  Figure~\ref{fig:ex_simus}) and we are able to visually associate some small spatial offsets values with a clear UV emission clump in the outskirt of the galaxy (one example shown in the top row of Figure~\ref{fig:ex_simus}). Due to the resolution limits, it is hard to distinguish the different potential UV emission clumps in the real observed LLAMAS galaxies, except for some highly magnified (less than 10 in the LLAMAS) objects as for example the source at $z=4.03$ in Figure~\ref{fig:offsets_im}. \\

Moreover we measured a significant ($>2$ HST pixels) distance between the UV brighter pixel (i.e. the UV peak emission location) and the UV emission centroids in 27\% of the LLAMAS sources (distribution shown in grey  in Figure~\ref{fig:off2_simus}) which reveals the clumpy nature of some galaxies and could explain the formation of some small offsets. Among these galaxies, 54\% present a spatial offset with $\Delta_{\rm ell}<1$, the presence of a clumpy UV emission distribution seems to favour the measurement of an internal offset in the galaxy. We notice the same trend in the simulated LLAMAS-like galaxies: 22\% of the objects present a significant UV-UV offset and among them 90\% present a value of $\Delta_{\rm ell}<1$. \\

On the other hand, 62 \% \ of the LLAMAS galaxies have an elliptical distance too high to be explained by internal substructures of the UV emission. Many other scenarios are compatible with these large offsets such as inflows of gas, emission from faint satellite galaxies, outflows in the CGM or Lyman-$\alpha$ scattering effects. In the LLAMAS galaxies, without new observations of ISM lines (such as {\sc H$\alpha$} or {[\oiii]}) or deepest UV data, it is difficult to disentangle these scenarios in each individual galaxy. The possibility of bright Lyman-$\alpha$ emission from faint satellite or merger galaxies was suggested by \citet{Maiolino2015} and \citet{Mitchell2021} and could explain both the larger values of ${r_{50,Ly\alpha}}/{r_{50,\rm UV}}$ and elliptical distances. Nevertheless, we did not clearly detect UV satellites coincident with this Lyman-$\alpha$ emission, even in the deepest HST fields. In the original simulation images, we can visually assign the spatial peak of the Lyman-$\alpha$ to a faint UV emission component located outside the main UV component, as shown in Figure~\ref{fig:ex_simus}, in 24\% of the LLAMAS simulated galaxies which represent the larger spatial offsets. 
The small measured offsets  ($<10$ kpc) suggest mainly cases of merger galaxies emission both in LLAMA and LLAMAS-like samples. 
We do not notice any significant trend between the spatial offset values and the physical properties of the LLAMAS-like simulated galaxies.

\subsubsection{Other possible origins and future direction}

%\subsection{Origin of spatial offsets in the lensed LAEs}

By comparing lensed LAE observations and a zoom-in simulation, we identified two different ranges of $\Delta$ values (cf. Figure~\ref{fig:r_ell}) showing that different scenarios are at play in the formation of spatial offsets between UV and Lyman-$\alpha$ emission at $z>3$:
\begin{enumerate}
\item Formation of small-scale offsets by bright UV clumps inside or in the outskirts of the main UV component. If most of the Lyman-$\alpha$ photons are produced in the ISM, the initial Lyman-$\alpha$ emission distribution should follow the ISM spatial morphology. In their simulated galaxy, \citet{Mitchell2021} showed that at $r<7$ kpc the ISM contribution dominate the Lyman-$\alpha$ emission. This scenario is coherent with LARS galaxies (\citealt{Ostlin2014}) at low redshift which are highly irregular and clumpy. The LARS study shows also that the Lyman-$\alpha$ $W_0$ is varying significantly as a function of the position in the galaxy. Recently, \citet{Armin2021} has measured the same type of spatial offsets between UV and Lyman-$\alpha$ emission centroids in low redshift galaxies (with offsets ranging from 0.8 to 2.25 kpc and a median value of $1.13$ kpc) and correlated to the stellar mass and the size of the star-formation regions. \citet{Guaita2015} showed that the UV component of these galaxies presents a morphology compatible with mergers or star-burst galaxies at high redshifts. \citet{Messa2019} measured that the LARS galaxies with a Lyman-$\alpha$ escape fraction higher than 10\% \ have more than 50\% \ of their UV luminosity which comes from UV stellar clumps. Finally, the turbulence in actively star-forming galaxies is strongly connected to ISM conditions that favour an escape of Lyman-$\alpha$ radiation (\citealt{Herenz2016, Kimm2019}).
In this  scenario, the offset could be intrinsic or affected by dust effects (\citealt{Behrens2019}), which could locally obstruct the emission of Lyman-$\alpha$ photons and thus produce a small spatial offset. In their sample, \citet{Hoag2019} noticed that the less dusty galaxies present on average a larger spatial offset. We do not report a clear trend between the UV slope $\beta$ and the spatial offset in the LLAMAS galaxies but we observed that the galaxies with larger offsets present higher $\beta$. Finally, the Lyman-$\alpha$ photons are produced in the vicinity of young short-lived stars while the $1500-$\AA \ UV emission arises on longer timescales and could be dominated by more evolved massive stars which produce less Lyman-$\alpha$ photons. An external young SFR cluster in the outskirts of the main UV component will therefore produce a spatial offset with the UV total light centroid and a spatial offset between UV and Lyman-$\alpha$ emission.

\item Lyman-$\alpha$ emission from faint UV satellites producing larger offsets values. This scenario was already proposed to explain either the formation of very extended Lyman-$\alpha$ haloes \citep{Mitchell2021}, or spatial offsets at $z \sim 3-7$ \citep{Shibuya2014,Maiolino2015}. \citet{Lemaux2020} measured a correlation between the UV brightness and the spatial offsets in kiloparsec as we measured in the LLAMAS galaxies with the SFR. The UV brightest galaxies are also the most massive, they are therefore more likely to reside in a more massive dark matter halo and thus be surrounded by faint satellite or merger galaxies.

\item Scattering effects of the Lyman-$\alpha$ photons across an in-homogeneous medium in such a way that Lyman-$\alpha$ emission seems to be offseted from the UV counterpart. This is more likely for the small offsets observed ($\Delta_{\rm ell}<2$). However, as the brightness of scattered Lyman-$\alpha$ emission decreases as a function of $1/r^{2}$, scattering effects alone are unlikely to produce the largest spatial offset such as those measured in the LLAMAS galaxies. The presence of ionised or low column density channels in the ISM and CGM, produced for instance by stellar feedbacks (\citealt{Rivera2017, Erb2019,Reddy2021}), could also produce this type spatial offsets.

\end{enumerate}

Each offset measured can also be produced by a combination of several of these phenomena, as proposed by \citet{Matthee2020_2} to explain a spatial offset measured in a $z=6.6$ galaxy. Another way to try to distinguish all these scenarios is to study the spatially resolved properties of the lines for the most extended objects. For example, \citet{Erb2019} measured in a spatial offset of 650 pc between UV and Lyman-$\alpha$ emission in a lensed galaxy at $z=1.84$ extended by 12 arcseconds. They explained this offset by a significant variation of the neutral hydrogen column density across the object, which supports a model in which ionizing radiation escapes from galaxies through channels with low column density of neutral gas. In a similar way, \citet{Chen2021M1206} identified 2 Lyman-$\alpha$ nebulae spatially offset from the associated star-forming regions. The variation of the Lyman-$\alpha$ surface brightness suggests large spatial fluctuations in the gas properties, and their results on spatial variations of the Lyman-$\alpha$ line profile, support a scenario in which high column density gas is driven toward up to 10 kpc. They conclude that the Lyman-$\alpha$ photons  originate from a combination of resonant scattering from the star-forming regions and recombination radiation due to escaping ionizing photons, but they were unable to determine the relative contribution of these two mechanisms.
A detailed study of the spectral and spatial properties of most extended Lyman-$\alpha$ haloes, as performed in \citet{Leclercq2020,Smit2017,Claeyssens2019} will allow us to detect potential variations of the CGM gas properties, such as hydrogen column density and kinematics, across the halo. In \citet{Claeyssens2019}, we studied the spatial variation of the Lyman-$\alpha$ line within  a lensed  halo at sub-kpc scale. We identified a region, in the outskirts of the halo, presenting a smaller spectral Lyman-$\alpha$ line shift (with respect to the systemic redshift) than the rest of the extended emission. The local  emission of Lyman-$\alpha$ photons by a faint, non-detected, UV component could explain the presence of weakly scattered Lyman-$\alpha$ photons at this location.

%--------------------------------------------------------------------
\section{Summary and conclusions}
\label{sec:conclusion}

We presented the largest statistical sample of lensed Lyman-$\alpha$ emitters observed with MUSE and HST. We observed 603 sources (producing 959 images) lensed by 17 different massive clusters.
Thanks to the lensing magnification (ranging from $1.4$ to $40$ in the total sample), we characterized the spatial properties of this new population of small and faint LAEs. We observe that $97$\% of the LLAMAS galaxies present an extended Lyman-$\alpha$ halo. We measured that the spatial extent of Lyman-$\alpha$ emission seem to be correlated with the UV SFR and the FWHM of the line.
We confirmed the correlation from \citet{Leclercq2017} between Lyman-$\alpha$ and UV emission spatial extent and extended it to fainter  LAEs but with a higher dispersion. We measured also the axis ratio of the UV continuum and Lyman-$\alpha$ emission distribution and notice that the $48$\% of the halo present an elliptical morphology (this fraction increase if we consider only the most resolved haloes). The Lyman-$\alpha$ haloes are on average less elliptical than the UV emission.
We measured secure spatial offsets between the UV and Lyman-$\alpha$ emissions for $63$\% of the sources, and found a distribution of values in very good agreement with \citet{Hoag2019, Lemaux2020, Ribeiro2020} with a median value of $\Delta=0.58$ kpc. We found very small or no correlations between the offset measurements and the physical parameters of the host galaxies (UV star formation rate, Lyman-$\alpha$ equivalent width and Lyman-$\alpha$ line FWHM). 
We identified 2 regimes in the offset distribution. First, galaxies with small offsets values ($38$\%) with respect to the UV emission distribution, are  likely due to bright star formation clumps in the outskirts of the UV component, emitting a strong Lyman-$\alpha$ emission. For the $62$\% other sources showing larger offsets, many scenarios could explain the large offsets such as inflows of gas, scattering effects of the photons in the CGM, extinction, outflows or Lyman-$\alpha$ emission from faint satellite galaxies not detected in UV. This last scenario is supported by the fact that we found higher values of ${r_{\rm Ly\alpha}}/{r_{\rm UV}}$ for galaxies with higher elliptical distances. 
Finally we compared our results with a zoom-in RHD simulation, following one typical faint Lyman-$\alpha$ emitter from $z=6$ to $z=3$, by producing, both in UV and Lyman-$\alpha$ emission, high resolution, "UDF-like" and "LLAMAS-like" mocks. The simulated galaxy is representative of the LLAMAS sample in terms of UV magnitude and Lyman-$\alpha$ halo size. We measured a similar spatial offsets distribution for the 3 samples and the LLAMAS galaxies. The simulation favors the interpretation where substructures in star-forming galaxies account for the smaller offsets (with $\Delta_{\rm ell}<2$) and satellites/merger galaxies explain the larger offsets (with $\Delta_{\rm ell}>2$). The scattering of Lyman-$\alpha$ photons could also contribute to production of spatial offsets.
 \\

It is likely that future works on these galaxies, especially the study of the spatial variations of emission lines profiles, and future observations of lower redshift LAEs with BlueMUSE \citep{Richard2019}, will help us understand more about the nature and the origin of the spatial offsets and Lyman-$\alpha$ haloes.

\begin{acknowledgements}
AC and JR acknowledge support from the ERC starting grant 336736-CALENDS. TG and AV are supported by the ERC Starting grant 757258 ‘TRIPLE’. HK, FL and AV acknowledge support from SNF Professorship PP00P2\_176808. FEB acknowledges support from ANID-Chile Basal AFB-170002, FONDECYT Regular 1200495 and 1190818, and Millennium Science Initiative Program  – ICN12\_009. HK acknowledges support from Japan Society for the Promotion of Science (JSPS) Overseas Research Fellowship. Tran Thi Thai was funded by Vingroup JSC and supported by the Master, PhD Scholarship Programme of Vingroup Innovation Foundation (VINIF), Institute of Big Data, code VINIF.2021.TS.041.
\end{acknowledgements}


\end{document}